\documentclass[floatfix,lengthcheck,showpacs,amssymb,amsmath,amsfonts,twocolumn,nofootinbib,longbibliography]{revtex4-1}
\usepackage{amsfonts,amsmath,units,wasysym,epsfig,graphicx,verbatim,color,subfigure,graphicx}
\usepackage{amsmath}
\usepackage{latexsym}
\usepackage{amssymb}
\usepackage{amsfonts}
\usepackage{mathtools}
\usepackage{bm}
\usepackage{color}
\usepackage{float}
\usepackage{tikz}
\usepackage{adjustbox}
\usepackage{array}
\usepackage{soul}
\usepackage{appendix}
\usepackage{physics}
\usepackage{braket}
\usepackage{xcolor}
\usepackage[none]{hyphenat}
\usepackage{lineno}

\newcommand{\approximant}[1]{{\fontfamily{qcr}\selectfont{#1} }}

\def\be{\begin{equation}}
\def\ee{\end{equation}}
\def\bea{\begin{eqnarray}}
\def\eea{\end{eqnarray}}

\def \D{{\cal D}}
\def \G{{\cal G}}

\def \S{{\cal S}}
\def \P{{\cal P}}
\def \N{{\cal N}}

\def \A{{\mathcal A}}

\def \a{\alpha}
\def \b{\beta}
\def \M{{\mathcal M}}

\def \x{{\bf x}}
\def \y{{\bf y}}
\def \h{{\bf h}}
\def \n{{\bf n}}
\def \g{{\bf g}}
\def \s{{\bf s}}

\def \e{{\bf e}}
\def \v{{\bf v}}
\def \w{{\bf w}}

\begin{document}

\author{\href{https://orcid.org/0000-0002-5077-8916}{Rahul Dhurkunde}} 
\affiliation{Max-Planck-Institut fur Gravitationsphysik (Albert-Einstein-Institut), D-30167 Hannover, Germany}
\affiliation{Leibniz Universitat Hannover, D-30167 Hannover, Germany}
\affiliation{Institute of Cosmology and Gravitation, University of Portsmouth, Portsmouth, PO1 3FX, U.K.}
\author{\href{https://orcid.org/0000-0002-1850-4587}{Alexander H. Nitz}}
\affiliation{Department of Physics, Syracuse University, Syracuse, NY 13244, USA}

\date{\today}
\title{Search for eccentric NSBH and BNS mergers in the third observing run of Advanced LIGO and Virgo}
 
\begin{abstract}
The possible formation histories of neutron star binaries remain unresolved by current gravitational-wave catalogs. The detection of an eccentric binary system could be vital in constraining compact binary formation models. We present the first search for aligned spin eccentric neutron star-black hole binaries (NSBH) and the most sensitive search for aligned-spin eccentric binary neutron star (BNS) systems using data from the third observing run of the advanced LIGO and advanced Virgo detectors. No new statistically significant candidates are found; we constrain the local merger rate for specific astrophysical models to be less than 150 $\text{Gpc}^{-3}\text{yr}^{-1}$ for binary neutron stars in the field, and, 50, 100, and 70 $\text{Gpc}^{-3}\text{yr}^{-1}$ for neutron star-black hole binaries in globular clusters, hierarchical triples and nuclear clusters, respectively, at the 90$\%$ confidence level if we assume that no sources have been observed from these populations. We predict the capabilities of upcoming and next-generation observatory networks; we investigate the ability of three LIGO ($\text{A}^{\#}$) detectors and Cosmic Explorer CE (20km) + CE (40km) to use eccentric binary observations for determining the formation history of neutron star binaries. We find that 2 -- 100 years of observation with three $\text{A}^{\#}$ observatories are required before we observe clearly eccentric NSBH binaries; this reduces to only 10 days -- 1 year with the CE detector network. CE will 
observe tens to hundreds of measurably eccentric binaries from
each of the formation models we consider.
\end{abstract}
 \maketitle

\section{Introduction}
Gravitational-wave (GW) astronomy is becoming routine; nearly $100$ compact binary mergers have been observed to date \cite{KAGRA:2021vkt,Nitz:2021zwj,Olsen:2022pin} using the Advanced LIGO  \cite{LIGOScientific:2014pky} and Advanced Virgo \cite{Acernese:2015gua} observatories. These observations have fueled interest in the long-standing question in astrophysics: \textit{how do compact binary systems form and evolve?} One class of models suggests these systems evolve as isolated stars in the field via common envelope \cite{Bethe:1998bn, Ivanova:2011wy}, stable mass transfer \cite{Soberman:1997mq} or via chemically homogeneous mixing \cite{Mandel:2015qlu}. Alternatively, they may be a result of a dynamical encounter of two or more separately evolved compact objects in dense environments such as globular clusters \cite{Ivanova:2007bu}, nuclear star clusters \cite{Fragione:2018yrb,Fragione:2019vgr,Neumayer:2020gno}, young star clusters \cite{Santoliquido:2020bry, Rastello:2020sru}, or active galactic nuclei \cite{McKernan:2020lgr}, (see \cite{Mandel:2021smh} for an overall review of the various channels). 
Current GW catalogs suggest that multiple formation pathways contribute to the population of binary black hole (BBH) mergers in the Universe rather than a single preferred channel \cite{Zevin:2020gbd,KAGRA:2021duu}. Pulsar observations indicate multiple formation channels for neutron star binaries (BNS or NSBH) \cite{Tauris:2017omb, Bernadich:2023uru}, however, the fewer GW observations of neutron star binaries is insufficient to determine if there is a preference for a single dominant channel or several competing channels present \cite{Andrews:2019vou,KAGRA:2021duu}.



Each formation channel makes distinct predictions for the distribution of binary properties, e.g. masses, spins, eccentricity and merger rate~\cite{Mandel:2021smh}. Distinguishing these channels could be done by careful comparison of large number of detected events or by identifying rare events with properties that are unique to a specific channel \cite{Zevin:2020gbd,Zevin:2021rtf}. Orbital eccentricity carries a strong signature of a binary's evolutionary history. In field binaries, energy dissipation solely occurs through GW emission, resulting in a swift reduction in eccentricity as the system evolves in frequency -- becoming nearly negligible when GW frequency reaches the sensitive band of current GW observatories (e.g 10 Hz) \cite{Peters:1964zz}. Whereas, in dense environments, angular momentum exchanges with a third compact object via the Lidov-Kozai (LK) mechanism \cite{LIDOV1962719,Kozai:1962zz, Antognini:2015loa} can result in sustained non-negligible eccentricities at GW frequencies sensitive to current detectors.

\begin{figure*}[]
    \centering
    \includegraphics[width=\linewidth, height=6.7cm]{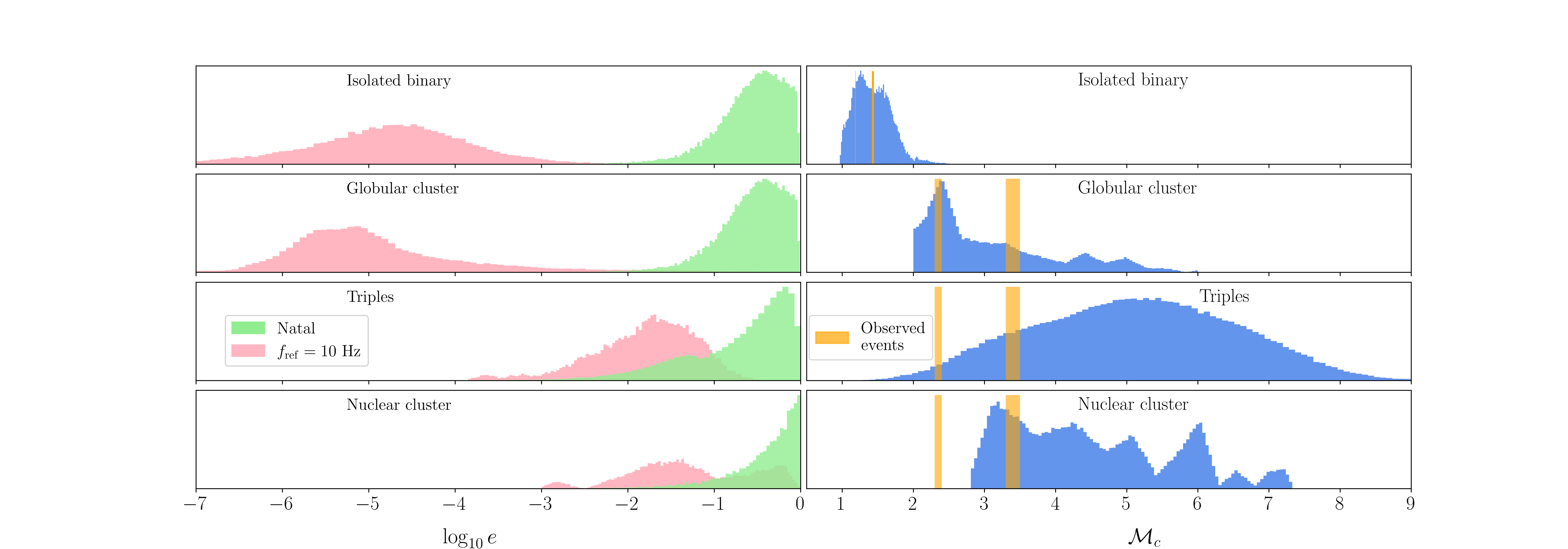}
    \caption{Distribution of orbital eccentricities (left column) for different formation models \cite{Belczynski:2001uc, Belczynski:2017mqx,Sedda:2020wzl,Trani:2021tan,Fragione:2018yrb, Fragione:2019vgr} at the time of formation of compact binaries (green) and when the dominant-mode of GW frequency reach 10 Hz (pink). All four models predict mergers to born with high natal eccentricities. Considering energy loss via GW only, eccentricity is quickly radiated away as evident by the clear shift in the distributions for the isolated BNS channel \cite{Belczynski:2001uc, Belczynski:2017mqx} and for NSBH mergers in globular clusters \cite{Sedda:2020wzl} -- natal eccentricity for BNS are evolved to 10 Hz using Peter's equation \cite{Peters:1964zz}. The nuclear cluster \cite{Fragione:2018yrb, Fragione:2019vgr} and hierarchical triples \cite{Trani:2021tan} models describe eccentricity enhancing scenarios where NSBH binaries can retain non-negligible eccentricities at 10 Hz -- eccentricity distribution of NSBHs in nuclear clusters \cite{Fragione:2018yrb, Fragione:2019vgr} is considered from triple systems that is predicted in \cite{Fragione:2019vgr}. We also show the chirp mass distribution predicted by each model in the second column and the estimated chirp masses of the observed BNS (GW170817, GW190425) and NSBH (GW200105, GW200115) events. We obtained chirp masses for NSBH sources in hierarchical triples and for isolated BNS systems assuming no correlation between the primary and secondary mass distributions. Chirp mass distribution for NSBH in globular clusters is the same as the histogram in the second row of Fig.5 in \cite{Sedda:2020wzl}.}
    \label{fig:eccentricity-dist}    
\end{figure*}

We highlight four formation scenarios in Fig. \ref{fig:eccentricity-dist} as a fiducial comparison \cite{Belczynski:2001uc,Belczynski:2017mqx,Sedda:2020wzl,Trani:2021tan,Fragione:2018yrb,Fragione:2019vgr}. We have considered two models without eccentricity-enhancing (LK) mechanisms -- one for isolated BNS binaries \cite{Belczynski:2001uc,Belczynski:2017mqx} and one for NSBH systems in globular clusters (strong hyperbolic interactions -- only restricted to a single exchange of a binary component with a third object) \cite{Sedda:2020wzl}. We also study two models with eccentricity-inducing (LK) mechanisms -- NSBH systems in hierarchical triples \cite{Trani:2021tan} and in nuclear clusters \cite{Fragione:2018yrb,Fragione:2019vgr}. In models influenced by the LK mechanism, up to $80\%$ of the systems could possess eccentricity $e_{10} \geq 0.01$ (eccentricity at dominant-mode gravitational-wave frequency of 10 Hz ($e_{10}$)) \cite{Hamers:2019oeq, Fragione:2018yrb, Fragione:2019vgr,Trani:2021tan, Silsbee:2016djf,Rodriguez:2018jqu}, and in the absence, only up to $5\%$ of the sources exceed this eccentricity \cite{Sedda:2020wzl, Belczynski:2001uc,Belczynski:2017mqx}. Observation of an eccentric system would clearly indicate the presence of a dynamical channel. Even a null detection would allow us to put tighter constraints on the predicted merger rates which are highly sensitive to the unconstrained parameters describing physical process such as common-envelope evolution \cite{Kowalska:2010qg}, natal supernovae kicks \cite{Chaurasia:2005aq, Richards:2022fnq} or dynamics of dense environments \cite{Fragione:2018yrb,Fragione:2019vgr,Trani:2021tan, Sedda:2020wzl}.

\begin{figure}[]
    \centering
    \includegraphics[width=\linewidth, height=6.8cm]{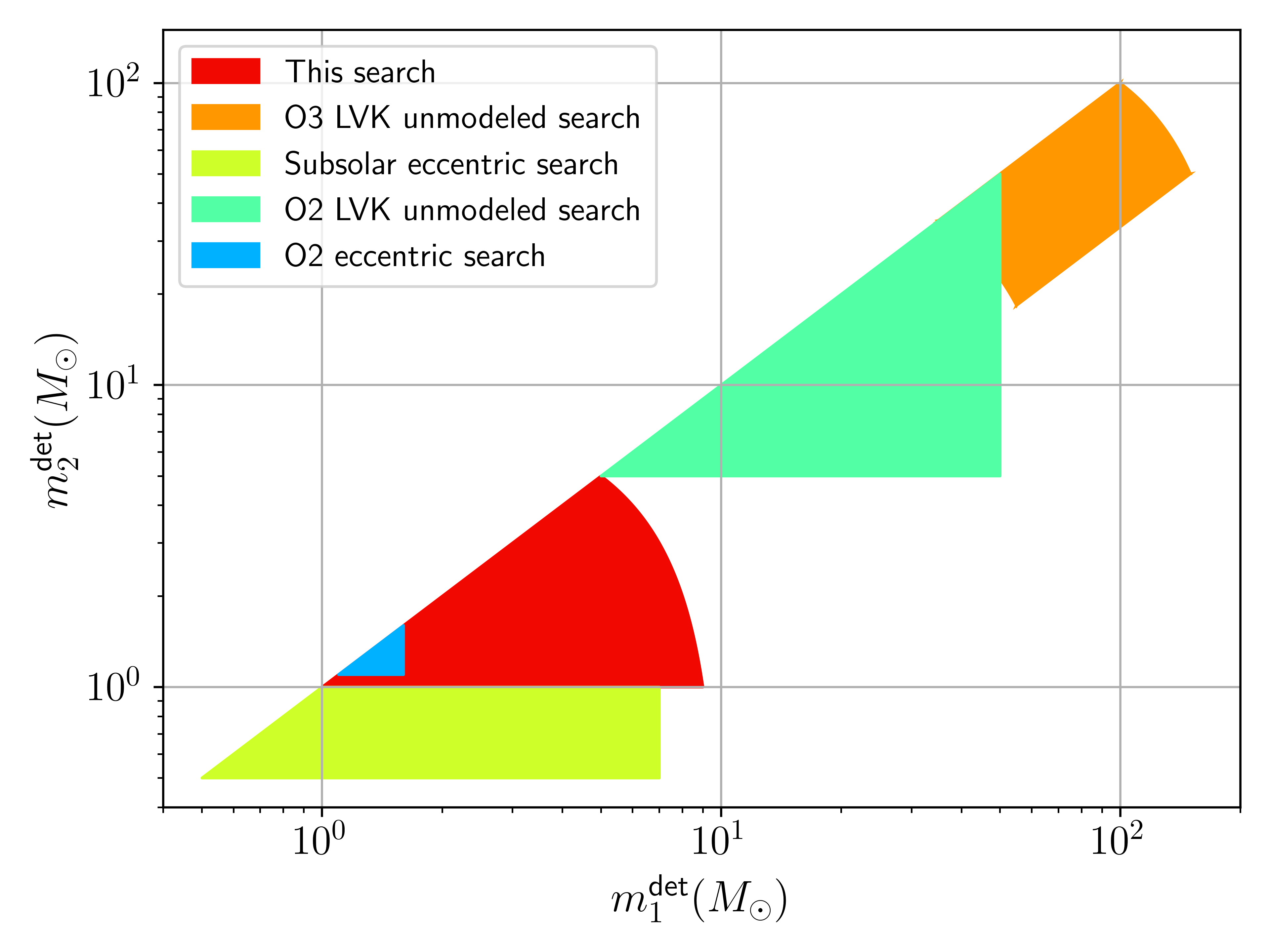}
    \caption{ Target regions of the various eccentric searches performed to date \cite{Nitz:2019spj,Nitz:2021vqh,LIGOScientific:2019dag,LIGOScientific:2023lpe} as a function of detector-frame masses ($m_1^{\text{det}}-m_2^{\text{det}}$). No prior searches have been explored for NSBH systems or BNS systems with spins. The prior search for BNS systems was restricted to a narrower region of masses and eccentricities ($e_{10} \leq 0.43$) and did not include spins \cite{Nitz:2019spj}. We search for spin-aligned neutron star binaries (BNS + NSBH) with eccentricities $e_{10} \leq 0.46$. The only searches for BBH sytems are unmodeled searches \cite{LIGOScientific:2019dag,LIGOScientific:2023lpe} and show the regions used to report their upper limits. The previous search for non spinning subsolar binaries restricted the eccentricities to $e_{10} \leq 0.3$ \cite{Nitz:2021vqh}.}%
    \label{fig:search_region}
\end{figure}

Four neutron star binary mergers have been observed till date: one BNS merger, GW170817 \cite{LIGOScientific:2017vwq}; a merger whose masses are consistent with being a BNS, GW190425 \cite{LIGOScientific:2020aai}; and two potential NSBH mergers GW200105 and GW200115 \cite{LIGOScientific:2021qlt}. All of these binaries were found using searches that only model quasi-circular binary orbits \cite{Usman:2015kfa, Messick:2016aqy, Aubin:2020goo,Chu:2020pjv}. If neutron star binaries have sufficiently high eccentricities, they would be missed by searches with Advanced LIGO data \cite{Huerta:2013qb}. The two BNS mergers have eccentricities limited to $e_{10} \leq 0.024$ and $e_{10} \leq 0.048$ for GW170817 and GW190425, respectively \cite{Lenon:2020oza}. A recent reanalysis of GW200105 has shown mild signs of eccentricity with $e_{20} = 0.145^{+0.007}_{-0.097}$ (90\% credible intervals) \cite{Morras:2025xfu}. In this Letter, we report the results for the first search for NSBH and the latest from BNS aligned-spin eccentric systems. A previous search for BNS mergers in the data from Advanced LIGO and Virgo's second observing run used a narrower range of binary masses (shown in Fig. \ref{fig:search_region}) and did not account component-object spins \cite{Nitz:2019spj}. Searches for eccentric subsolar binaries have also been performed \cite{Nitz:2021vqh, Nitz:2021mzz}. Unmodeled ~\cite{LIGOScientific:2019dag, LIGOScientific:2023lpe} and modeled \cite{Pal:2023dyg} searches have been performed for eccentric stellar-mass BBH systems. While these searches did not yield any new candidates, they constrained the local merger rate to be less than: $1700$ mergers $\text{Gpc}^{-3} \text{yr}^{-1}$ for BNS systems with $e_{10} \leq 0.43$ and $0.33$ mergers $\text{Gpc}^{-3} \text{yr}^{-1}$ for BBH systems with total mass $M \in [70M_{\odot},200M_{\odot}]$ and $e_{15} < 0.3$ at 90\% confidence. In addition, radio pulsar surveys have discovered over twenty BNS systems exhibiting a broad range of eccentricities between 0.06 and 0.8 at the very early inspiral stage \cite{Bernadich:2023uru} (see Table I). These observations constrain the BNS local merger rate to $293^{+222}_{-103}$ $\text{Gpc}^{-3} \text{yr}^{-1}$.

We do not find any new mergers in the public data from the third observing run (O3) of Advanced LIGO and Advanced Virgo observatories. We use our observations and the capabilities of future observatories to constrain an isolated BNS model and three different models for NSBH mergers in globular clusters, nuclear clusters and hierarchical triples: our observations restrict the rate of mergers for BNS binaries to be less than $\sim 150$ mergers $\text{Gpc}^{-3}\text{yr}^{-1}$ and less than $\sim $ 100 mergers $\text{Gpc}^{-3}\text{yr}^{-1}$ for NSBH binaries at 90\% confidence. These constraints assume that the prior observed BNS and NSBH mergers are from alternate formation channels. Assuming they are from one or more of the channels we consider, the measured rate would be consistent given the sparsity of observations. We predict the capabilities of improved second generation and upcoming third-generation GW observatories to use eccentric binary observations to constraint formation models. We find that a network of Cosmic Explorer (CE) \cite{Evans:2021gyd} (40 km) + CE (20km) observatories will detect the majority of sources from each of these models and could determine that a subset of the population have non-negligible eccentricities. A network of three $\text{A}^{\#}$ observatories 
would require at least $\sim$ two years of observation to detect a non-negligible eccentric NSBH merger from hierarchical triples or nuclear clusters.    

\section{Search description and observational results}
To search for eccentric binaries, we use the PyCBC toolkit to perform a template-based matched filtering analysis to find modeled GW signals in the interferometric data \cite{Usman:2015kfa,Nitz:2017svb}. GW candidates are identified by finding peaks of the signal-to-noise (SNR) time-series, mitigating non-Gaussian noise artefacts, and checking the consistency of the data and astrophysical sources between each detector~\cite{Allen:2004gu, Nitz:2017lco,Davies:2020tsx}. Taking into account these factors and the empirically measured noise distribution, each candidate is assigned a ranking statistic value \cite{Nitz:2017svb, Davies:2020tsx, Was:2009vh}.

We search for neutron star binaries using a bank of modeled waveforms (templates) generated using a stochastic placement method \cite{Harry:2009ea,Babak:2008rb}. Our search region is described by five binary parameters: detector-frame component masses ($m_1^{\text{det}}, m_2^{\text{det}}$) ranging from [1.0, 9.0] $M_{\odot}$ with cutoff on the total mass $M \leq 10 M_{\odot}$, $z-$ component of the individual spins ($|s_{1z}|, |s_{2z}| \in [0.0, 0.1]$), eccentricity $e_{20}$ at 20 Hz $\in [0, 0.28]$, and an additional source orientation parameter $l$ related to the position of the periapsis. Eccentricities in our template bank are defined at the lowest frequency used in our analysis which are then converted to the standard $f_{\text{ref}}$ = 10 Hz using the same eccentricity evolution as \approximant{TaylorF2Ecc} (described below) to analyze the various population models. Our eccentric bank contains $\sim$ 6 million templates which is roughly two orders of magnitude larger than an equivalent bank for quasi-circular binaries. To model the GW signals, we use 
\approximant{TaylorF2Ecc} inspiral only waveform model \cite{Moore:2016qxz} (from LALSuite \cite{lalsuite}) which accounts for non-spinning eccentric corrections to the quasi-circular \approximant{TaylorF2} model \cite{Bohe:2013cla}. \approximant{TaylorF2Ecc} models the spin-spin coupling up to 3PN \cite{Bohe:2015ana} and BH spin induced quadrupoles with no tidal effects.  The search is reliable when using just the inspiral segment of a signal, as only the inspiral part contributes dominantly to a signal's SNR. 


We search the O3 public Advanced LIGO and Virgo datasets using broadly the same search methods as \cite{Nitz:2021zwj}. O3 was divided into two parts -- O3a and O3b, comprising in total of $\sim 272$ days of coincident time when at least two observatories were in operation \cite{KAGRA:2023pio}. Our search did not find any new significant GW candidates. We recovered the previously reported multi-detector NSBH event GW200115 with high significance. As anticipated, we missed GW190425 and GW200105 since they were detected by a single detector. The most significant candidate has a FAR of about 1 per year, consistent with the null hypothesis based on the observation duration. The list of top candidates, the template parameters associated with each candidate, and the configuration files necessary to reproduce the analysis are available in our data release~\cite{github}.  

\section{Constraining population models}

For a given astrophysical model with a merger rate density $\mathcal{R}(\theta, z)$, the expected number of detections within an observation period $T_{\text{obs}}$ is 
\begin{align}
    N_{\text{detected}} & = T_{\text{obs}} \int \int \mathcal{R}(\theta,z) f(\theta,z) \dfrac{dV_c}{dz}\dfrac{1}{1+z} d\theta dz ,
    \label{Eq:expected-detections}
\end{align}
where $\theta$ is the set of various binary parameters predicted by an astrophysical model, $dV_c/dz$ is the differential co-moving volume and $f(\theta, z)$ is the probability of detecting a merger with $\theta$ parameters at a redshift $z$. We can constrain the local merger rate using the lack of observations: if we assume a Poisson distribution of observed mergers, then the 90\% confidence limit $R_{90}^{\text{local}}$ corresponds to a local merger rate when the expected number of detections is $\sim 2.3$ \cite{Biswas:2007ni}.

Upper limits are obtained by estimating the expected number of detections $N_{\text{detected}}$ (Eq. \ref{Eq:expected-detections}) via a Monte Carlo (MC) integration scheme for a synthetic population of mergers with binary parameter distributions predicted by the respective models \cite{Tiwari:2017ndi}. For a search, the detection probabilities can be estimated by using the search to detect simulated sources injected into the data. In our simulations, we assume the merger rate density follows the star formation rate \cite{Madau:2016jbv} convolved with the inverse time-delay distribution, same as the method used in \cite{Zhu:2020ffa, Wu:2022pyg}. The injection results from our search and the codes to estimate the observational limits are available as a part of our data release \cite{github}.

\begin{figure*}[]
    \centering
    \includegraphics[width=\textwidth, height=6.3cm]{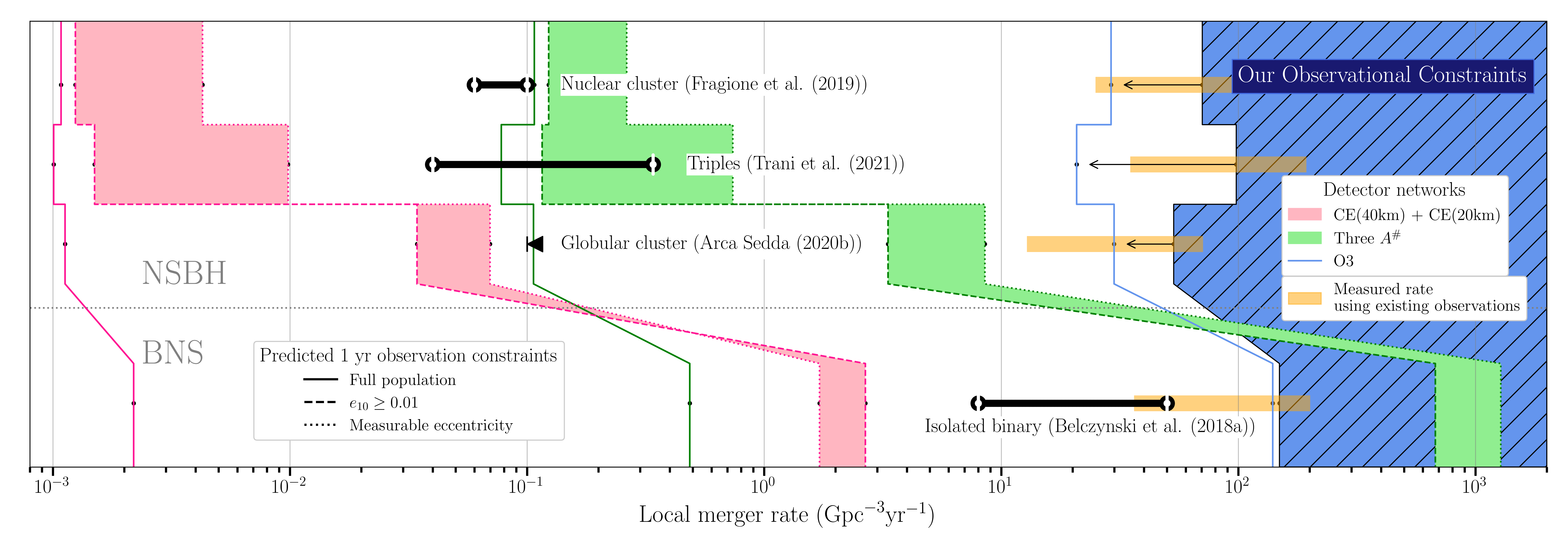}
    \caption{ Predicted and observed constraints on the local merger rate for various populations of BNS and NSBH sources \cite{Belczynski:2001uc,Belczynski:2017mqx, Sedda:2020wzl,Trani:2021tan,Fragione:2018yrb,Fragione:2019vgr}. Assuming the observed neutron star binaries are part of the channels we considered, our measured rate (horizontal orange bars) is consistent with the observed rates in the existing GW catalogs \cite{Nitz:2021zwj,KAGRA:2021vkt,Olsen:2022pin}. The observational constraints assuming a null detection from our search for 272 days of observation is shown as blue (hatched) region. The $y$-axis acts as a visual guide, enabling comparison of different model limits at specific heights. The upward transition from BNS to NSBH, represented by slanted lines, indicates a shift between distinct models and should not be interpreted as a continuous progression. Predicted constraints for an idealized search are shown for O3 (blue), three $\text{A}^{\#}$ (green) and CE (40km) + CE (20km) (pink) for an year of observation. In an idealized search, any merger from a given model can be detected if they exceed a network SNR of 10: achieving this, our search constraints would be up to $5\times$ tighter reaching the idealized O3 limits (solid blue line). The upper limits for systems with $e_{10}\geq0.01$ or with measurable nonzero eccentricity are shown as dashed and dotted lines respectively. These predictions allow us to estimate the time required for an eccentric merger observation; the hierarchical triples model predicts a maximum merger rate of 0.34 $\text{Gpc}^{-3}\text{yr}^{-1}$ and the $\text{A}^{\#}$ limit for systems with measurable eccentricity is 0.74 $\text{Gpc}^{-3}\text{yr}^{-1}$, this gives an expected 2.2 years of observations with $\text{A}^{\#}$ for an eccentric NSBH merger observation. To maintain readability of the plot, A+ and ET results are omitted from the figure but are fully available in our data release.} 
    \label{fig:time-requirement}    
\end{figure*}

\section{Astrophysical implications} 

We investigate how our observational results and the capability of future detectors can constraint four different astrophysical models: three dynamical pathways for NSBH systems within nuclear clusters \cite{Fragione:2018yrb,Fragione:2019vgr}, globular clusters \cite{Sedda:2020wzl}, and hierarchical triples \cite{Trani:2021tan}, and a BNS formation model in the field \cite{Belczynski:2001uc,Belczynski:2017mqx} to contrast the two major types of channels. We use the mass and eccentricity histograms provided in \cite{Belczynski:2001uc,Belczynski:2017mqx, Sedda:2020wzl,Trani:2021tan,Fragione:2018yrb,Fragione:2019vgr} and assume no correlation between the parameters. The predicted merger rates for the considered models are shown in Fig. \ref{fig:time-requirement}, and the rates from various other models can span up to five orders of magnitude due to the large uncertainties in these models \cite{Mandel:2021smh}. We constrain the local merger rate to be less than 150 $\text{Gpc}^{-3}\text{yr}^{-1}$ (isolated BNS), 50 $\text{Gpc}^{-3}\text{yr}^{-1}$ (NSBH in globular clusters), 100 $\text{Gpc}^{-3}\text{yr}^{-1}$ (NSBH in hierarchical triples) and 70 $\text{Gpc}^{-3}\text{yr}^{-1}$ (NSBH in nuclear clusters) under the assumption of non-detection from these channels. Our search region covers 99\% of BNS \cite{Belczynski:2001uc,Belczynski:2017mqx} and  85\%, 30\% and 45\% of the NSBH systems \cite{Sedda:2020wzl, Trani:2021tan, Fragione:2018yrb,Fragione:2019vgr}, respectively. Out of which the the current detectors can detect $5\%$ of BNS and up to $37\%$ of NSBH sources. We also measure the rate of neutron star binaries from these models, assuming all prior observations are associated with each channel; these are consistent due to the lack of observations to date. Clearly, current GW observatories cannot constrain the dynamical formation models we have considered. 

For each astrophysical model, we assume the spins to be aligned with ($|s_{1z}|, |s_{2z}| \in [0, 0.1]$): low spins are consistent with the current observations of galactic neutron star binaries (see the discussion in \cite{LIGOScientific:2018hze}). To assess the bias in observational constraints solely due to the restricted spins, we assume BH spins with spin magnitudes uniformly distributed as $ |s_{1z}| \in [0, 0.99]$ and distributed isotropically. We find using \approximant{IMRPhenomXPHM} \cite{Pratten:2020ceb,Pratten:2020fqn}, that neglecting higher-spin amplitudes, precession, and HMs have less than $13\%$ bias on the observational constraints on the local merger rate. Matter effects are insignificant for the population models we have considered -- tidal distruption frequencies exceed the maximum frequency used in our analysis, and the fitting-factor loss as reported in \cite{Cullen:2017oaz} remains negligible when averaged separately over each model. We investigate the potential waveform systematics on our chosen waveform model by comparing with \approximant{TEOBResumS} \cite{Nagar:2021xnh} and bound any potential observational bias to be less than $10\%$. However, waveform systematics can bias the predicted rates for $A^{\#}$ by up to 60\%, implying that improved waveform models will be required for analyses with $A^{\#}$ and beyond.

Improved second generation and upcoming third generation observatories are expected to be a factor of a few and more than an order of magnitude more sensitive than the current ones, respectively \cite{Evans:2021gyd, Srivastava:2022slt, post-o5-curves}. Third generation detectors will be sensitive to the majority of neutron star binaries in the Universe. So the question arises: \textit{ To what extent can future observatories determine the formation history of neutron star binaries?} We predict how well upcoming second and third generation observatories will be able to constrain these models using an idealized search capturing the full inspiral-merger-ringdown SNR by using the \approximant{IMRPhenomD} model \cite{Husa:2015iqa, Khan:2015jqa}. We investigate constraints on the local merger rate for two networks (shown in Fig. \ref{fig:time-requirement}) -- one consisting of three $\text{A}^{\#}$ \cite{post-o5-curves, Asharp_sensitivity} observatories and another composed of CE (baseline 40km) + CE (baseline 20km) \cite{CE_sensitivity, Gupta:2023lga} using their expected noise curves \cite{Evans:2021gyd, Srivastava:2022slt}. Furthermore, we have constraints for networks involving A+ \cite{post-o5-curves-obs-paper, post-o5-curves} and/or Einstein Telescope (ET) \cite{et-noise-curve, Hild:2008ng, Hild:2009ns, Hild:2010id} which are not presented here but are available in our data release \cite{github}. In agreement with \cite{Baibhav:2019gxm}, we find that CE will be able to detect majority of sources from each model. While current detectors may require up to $\mathcal{O}(10^3)$ years of observation to observe mergers from the considered dynamical formation models, three $\text{A}^{\#}$ observatories would begin detecting events from the most optimistic of these channels in roughly two years, three A+ in at least 20 years, and CE (40km) + CE (20km) (and similarly ET) with a few days of observation.    
 
Even though we can observe binaries from various formation channels, only those with high eccentricities can be clearly differentiated from non-eccentric binaries. Future observatory observations might struggle to confidently attribute binaries to dynamical channels unless they exhibit high eccentricity. To elucidate this, we show the population constraints for a fixed eccentricity threshold of $e_{10} \geq 0.01$ in Fig.~\ref{fig:time-requirement}. The limits for a fixed threshold scales inversely to the predicted fraction of systems satisfying this threshold: limits for mergers with $e_{10} \geq 0.01$ for NSBH in hierarchical triples or nuclear cluster models are worse only by a few factor due to large fraction of such sources predicted (see Fig. \ref{fig:eccentricity-dist}). 

The ability to measure eccentricity depends on the properties of a binary and the capabilities of a detector network  \cite{Lower:2018seu}, which a fixed threshold cannot capture -- we assess the potential of future observatories to measure eccentricity for each binary in our simulated population models using only phase corrections due to orbital eccentricity. We use a constrained parameter space to account for mild correlation between $\mathcal{M}_c$ and $e_{10}$ parameters \cite{Favata:2021vhw}. We use a simplified likelihood and assume a zero noise realization \cite{Pankow:2018qpo} that maximizes over the extrinsic parameters based on the fitting factor \cite{Baird:2012cu}. We sample over the two-dimensional parameter space using MCMC, employing the Dynesty sampler \cite{Skilling:2006gxv, Higson2018}, to derive HPD credible intervals \cite{Chen1999}. Our simplified likelihood is consistent with the extensive parameter estimation results, and comparison plots are provided in our data release. We deem a source to have measurable eccentricity if, at the 90\% credible level, we can rule out the quasi-circular binary hypothesis. We find the threshold $e_{10} \geq 0.01$ falls short for measuring eccentric neutron star binaries with three $\text{A}^{\#}$ or for eccentric NSBH binaries with the CE network. In contrast, the CE network is more proficient in measuring eccentricities of BNS systems: the same threshold is overly pessimistic for BNS. Crucially, we find that third-generation observatories are poised to detect eccentric BNS systems even in isolated binary channels. Future analyses may require computational resources up to two orders of magnitude greater than current analyses, necessitating the development of efficient search techniques such as hierarchical searches to ensure feasibility \cite{Dhurkunde:2021csz, Soni:2023veu}.


Fig. \ref{fig:time-requirement} suggests that the CE detector network will observe hundreds of highly eccentric NSBH sources from nuclear clusters or hierarchical triples; their non-detection would require the channel to have lower merger rates, prompting tighter constraints on the model parameters. For example, in nuclear clusters, the distribution of stars around supermassive black holes (typically depicted by a power law $n(r) \propto r^{-\alpha}$) influences the eccentricity profile of NSBH systems \cite{Fragione:2018yrb,Fragione:2019vgr}. An increase in $\alpha$ corresponds to more eccentric systems, so a non detection of eccentric sources would constrain the distribution of stars. CE will also measure eccentricities of isolated BNS mergers; with a model of natal orbital separations, one could estimate the distribution of natal eccentricities. Natal eccentricities are highly sensitive to the supernovae kick velocity \cite{Hobbs:2005yx, Richards:2022fnq}, and their estimation would allow constraints on the kick velocity.

\acknowledgments 
We would like to acknowledge Xisco J. Forteza, Sumit Kumar and Shichao Wu for the helpful discussions and feedback on the manuscript. We acknowledge the Max Planck Gesellschaft and the Atlas cluster computing team at Albert-Einstein Institute (AEI) Hannover for support. AHN acknowledges support from NSF grant PHY-2309240. This research has made use of data or software obtained from the Gravitational Wave Open Science Center (gwosc.org), a service of the LIGO Scientific Collaboration, the Virgo Collaboration, and KAGRA. This material is based upon work supported by NSF's LIGO Laboratory which is a major facility fully funded by the National Science Foundation, as well as the Science and Technology Facilities Council (STFC) of the United Kingdom, the Max-Planck-Society (MPS), and the State of Niedersachsen/Germany for support of the construction of Advanced LIGO and construction and operation of the GEO600 detector. Additional support for Advanced LIGO was provided by the Australian Research Council. Virgo is funded, through the European Gravitational Observatory (EGO), by the French Centre National de Recherche Scientifique (CNRS), the Italian Istituto Nazionale di Fisica Nucleare (INFN) and the Dutch Nikhef, with contributions by institutions from Belgium, Germany, Greece, Hungary, Ireland, Japan, Monaco, Poland, Portugal, Spain. KAGRA is supported by Ministry of Education, Culture, Sports, Science and Technology (MEXT), Japan Society for the Promotion of Science (JSPS) in Japan; National Research Foundation (NRF) and Ministry of Science and ICT (MSIT) in Korea; Academia Sinica (AS) and National Science and Technology Council (NSTC) in Taiwan.

\clearpage
\bibliography{references}

\begin{thebibliography}{100}%
\makeatletter
\providecommand \@ifxundefined [1]{%
 \@ifx{#1\undefined}
}%
\providecommand \@ifnum [1]{%
 \ifnum #1\expandafter \@firstoftwo
 \else \expandafter \@secondoftwo
 \fi
}%
\providecommand \@ifx [1]{%
 \ifx #1\expandafter \@firstoftwo
 \else \expandafter \@secondoftwo
 \fi
}%
\providecommand \natexlab [1]{#1}%
\providecommand \enquote  [1]{``#1''}%
\providecommand \bibnamefont  [1]{#1}%
\providecommand \bibfnamefont [1]{#1}%
\providecommand \citenamefont [1]{#1}%
\providecommand \href@noop [0]{\@secondoftwo}%
\providecommand \href [0]{\begingroup \@sanitize@url \@href}%
\providecommand \@href[1]{\@@startlink{#1}\@@href}%
\providecommand \@@href[1]{\endgroup#1\@@endlink}%
\providecommand \@sanitize@url [0]{\catcode `\\12\catcode `\$12\catcode `\&12\catcode `\#12\catcode `\^12\catcode `\_12\catcode `\%12\relax}%
\providecommand \@@startlink[1]{}%
\providecommand \@@endlink[0]{}%
\providecommand \url  [0]{\begingroup\@sanitize@url \@url }%
\providecommand \@url [1]{\endgroup\@href {#1}{\urlprefix }}%
\providecommand \urlprefix  [0]{URL }%
\providecommand \Eprint [0]{\href }%
\providecommand \doibase [0]{http://dx.doi.org/}%
\providecommand \selectlanguage [0]{\@gobble}%
\providecommand \bibinfo  [0]{\@secondoftwo}%
\providecommand \bibfield  [0]{\@secondoftwo}%
\providecommand \translation [1]{[#1]}%
\providecommand \BibitemOpen [0]{}%
\providecommand \bibitemStop [0]{}%
\providecommand \bibitemNoStop [0]{.\EOS\space}%
\providecommand \EOS [0]{\spacefactor3000\relax}%
\providecommand \BibitemShut  [1]{\csname bibitem#1\endcsname}%
\let\auto@bib@innerbib\@empty
\bibitem [{\citenamefont {Abbott}\ \emph {et~al.}(2023{\natexlab{a}})\citenamefont {Abbott} \emph {et~al.}}]{KAGRA:2021vkt}%
  \BibitemOpen
  \bibfield  {author} {\bibinfo {author} {\bibfnamefont {R.}~\bibnamefont {Abbott}} \emph {et~al.} (\bibinfo {collaboration} {KAGRA, VIRGO, LIGO Scientific}),\ }\bibfield  {title} {\enquote {\bibinfo {title} {{GWTC-3: Compact Binary Coalescences Observed by LIGO and Virgo during the Second Part of the Third Observing Run}},}\ }\href {\doibase 10.1103/PhysRevX.13.041039} {\bibfield  {journal} {\bibinfo  {journal} {Phys. Rev. X}\ }\textbf {\bibinfo {volume} {13}},\ \bibinfo {pages} {041039} (\bibinfo {year} {2023}{\natexlab{a}})},\ \Eprint {http://arxiv.org/abs/2111.03606} {arXiv:2111.03606 [gr-qc]} \BibitemShut {NoStop}%
\bibitem [{\citenamefont {Nitz}\ \emph {et~al.}(2023)\citenamefont {Nitz}, \citenamefont {Kumar}, \citenamefont {Wang}, \citenamefont {Kastha}, \citenamefont {Wu}, \citenamefont {Sch\"afer}, \citenamefont {Dhurkunde},\ and\ \citenamefont {Capano}}]{Nitz:2021zwj}%
  \BibitemOpen
  \bibfield  {author} {\bibinfo {author} {\bibfnamefont {Alexander~H.}\ \bibnamefont {Nitz}}, \bibinfo {author} {\bibfnamefont {Sumit}\ \bibnamefont {Kumar}}, \bibinfo {author} {\bibfnamefont {Yi-Fan}\ \bibnamefont {Wang}}, \bibinfo {author} {\bibfnamefont {Shilpa}\ \bibnamefont {Kastha}}, \bibinfo {author} {\bibfnamefont {Shichao}\ \bibnamefont {Wu}}, \bibinfo {author} {\bibfnamefont {Marlin}\ \bibnamefont {Sch\"afer}}, \bibinfo {author} {\bibfnamefont {Rahul}\ \bibnamefont {Dhurkunde}}, \ and\ \bibinfo {author} {\bibfnamefont {Collin~D.}\ \bibnamefont {Capano}},\ }\bibfield  {title} {\enquote {\bibinfo {title} {{4-OGC: Catalog of Gravitational Waves from Compact Binary Mergers}},}\ }\href {\doibase 10.3847/1538-4357/aca591} {\bibfield  {journal} {\bibinfo  {journal} {Astrophys. J.}\ }\textbf {\bibinfo {volume} {946}},\ \bibinfo {pages} {59} (\bibinfo {year} {2023})},\ \Eprint {http://arxiv.org/abs/2112.06878} {arXiv:2112.06878 [astro-ph.HE]} \BibitemShut {NoStop}%
\bibitem [{\citenamefont {Olsen}\ \emph {et~al.}(2022)\citenamefont {Olsen}, \citenamefont {Venumadhav}, \citenamefont {Mushkin}, \citenamefont {Roulet}, \citenamefont {Zackay},\ and\ \citenamefont {Zaldarriaga}}]{Olsen:2022pin}%
  \BibitemOpen
  \bibfield  {author} {\bibinfo {author} {\bibfnamefont {Seth}\ \bibnamefont {Olsen}}, \bibinfo {author} {\bibfnamefont {Tejaswi}\ \bibnamefont {Venumadhav}}, \bibinfo {author} {\bibfnamefont {Jonathan}\ \bibnamefont {Mushkin}}, \bibinfo {author} {\bibfnamefont {Javier}\ \bibnamefont {Roulet}}, \bibinfo {author} {\bibfnamefont {Barak}\ \bibnamefont {Zackay}}, \ and\ \bibinfo {author} {\bibfnamefont {Matias}\ \bibnamefont {Zaldarriaga}},\ }\bibfield  {title} {\enquote {\bibinfo {title} {{New binary black hole mergers in the LIGO-Virgo O3a data}},}\ }\href {\doibase 10.1103/PhysRevD.106.043009} {\bibfield  {journal} {\bibinfo  {journal} {Phys. Rev. D}\ }\textbf {\bibinfo {volume} {106}},\ \bibinfo {pages} {043009} (\bibinfo {year} {2022})},\ \Eprint {http://arxiv.org/abs/2201.02252} {arXiv:2201.02252 [astro-ph.HE]} \BibitemShut {NoStop}%
\bibitem [{\citenamefont {Aasi}\ \emph {et~al.}(2015)\citenamefont {Aasi} \emph {et~al.}}]{LIGOScientific:2014pky}%
  \BibitemOpen
  \bibfield  {author} {\bibinfo {author} {\bibfnamefont {J.}~\bibnamefont {Aasi}} \emph {et~al.} (\bibinfo {collaboration} {LIGO Scientific}),\ }\bibfield  {title} {\enquote {\bibinfo {title} {{Advanced LIGO}},}\ }\href {\doibase 10.1088/0264-9381/32/7/074001} {\bibfield  {journal} {\bibinfo  {journal} {Class. Quant. Grav.}\ }\textbf {\bibinfo {volume} {32}},\ \bibinfo {pages} {074001} (\bibinfo {year} {2015})},\ \Eprint {http://arxiv.org/abs/1411.4547} {arXiv:1411.4547 [gr-qc]} \BibitemShut {NoStop}%
\bibitem [{\citenamefont {Acernese}(2015)}]{Acernese:2015gua}%
  \BibitemOpen
  \bibfield  {author} {\bibinfo {author} {\bibfnamefont {F.}~\bibnamefont {Acernese}} (\bibinfo {collaboration} {Virgo}),\ }\bibfield  {title} {\enquote {\bibinfo {title} {{The Advanced Virgo detector}},}\ }\href {\doibase 10.1088/1742-6596/610/1/012014} {\bibfield  {journal} {\bibinfo  {journal} {J. Phys. Conf. Ser.}\ }\textbf {\bibinfo {volume} {610}},\ \bibinfo {pages} {012014} (\bibinfo {year} {2015})}\BibitemShut {NoStop}%
\bibitem [{\citenamefont {Bethe}\ and\ \citenamefont {Brown}(1998)}]{Bethe:1998bn}%
  \BibitemOpen
  \bibfield  {author} {\bibinfo {author} {\bibfnamefont {Hans~A.}\ \bibnamefont {Bethe}}\ and\ \bibinfo {author} {\bibfnamefont {G.~E.}\ \bibnamefont {Brown}},\ }\bibfield  {title} {\enquote {\bibinfo {title} {{Evolution of binary compact objects which merge}},}\ }\href {\doibase 10.1086/306265} {\bibfield  {journal} {\bibinfo  {journal} {Astrophys. J.}\ }\textbf {\bibinfo {volume} {506}},\ \bibinfo {pages} {780--789} (\bibinfo {year} {1998})},\ \Eprint {http://arxiv.org/abs/astro-ph/9802084} {arXiv:astro-ph/9802084} \BibitemShut {NoStop}%
\bibitem [{\citenamefont {Ivanova}(2011)}]{Ivanova:2011wy}%
  \BibitemOpen
  \bibfield  {author} {\bibinfo {author} {\bibfnamefont {Natalia}\ \bibnamefont {Ivanova}},\ }\bibfield  {title} {\enquote {\bibinfo {title} {{Common envelope: the progress and the pitfalls}},}\ }\href@noop {} {\bibfield  {journal} {\bibinfo  {journal} {ASP Conf. Ser.}\ }\textbf {\bibinfo {volume} {447}},\ \bibinfo {pages} {91} (\bibinfo {year} {2011})},\ \Eprint {http://arxiv.org/abs/1108.1226} {arXiv:1108.1226 [astro-ph.SR]} \BibitemShut {NoStop}%
\bibitem [{\citenamefont {Soberman}\ \emph {et~al.}(1997)\citenamefont {Soberman}, \citenamefont {Phinney},\ and\ \citenamefont {den Heuvel}}]{Soberman:1997mq}%
  \BibitemOpen
  \bibfield  {author} {\bibinfo {author} {\bibfnamefont {G.~E.}\ \bibnamefont {Soberman}}, \bibinfo {author} {\bibfnamefont {E.~S.}\ \bibnamefont {Phinney}}, \ and\ \bibinfo {author} {\bibfnamefont {E.~P. J.~van}\ \bibnamefont {den Heuvel}},\ }\bibfield  {title} {\enquote {\bibinfo {title} {{Stability criteria for mass transfer in binary stellar evolution}},}\ }\href@noop {} {\bibfield  {journal} {\bibinfo  {journal} {Astron. Astrophys.}\ }\textbf {\bibinfo {volume} {327}},\ \bibinfo {pages} {620} (\bibinfo {year} {1997})},\ \Eprint {http://arxiv.org/abs/astro-ph/9703016} {arXiv:astro-ph/9703016} \BibitemShut {NoStop}%
\bibitem [{\citenamefont {Mandel}\ and\ \citenamefont {de~Mink}(2016)}]{Mandel:2015qlu}%
  \BibitemOpen
  \bibfield  {author} {\bibinfo {author} {\bibfnamefont {Ilya}\ \bibnamefont {Mandel}}\ and\ \bibinfo {author} {\bibfnamefont {Selma~E.}\ \bibnamefont {de~Mink}},\ }\bibfield  {title} {\enquote {\bibinfo {title} {{Merging binary black holes formed through chemically homogeneous evolution in short-period stellar binaries}},}\ }\href {\doibase 10.1093/mnras/stw379} {\bibfield  {journal} {\bibinfo  {journal} {Mon. Not. Roy. Astron. Soc.}\ }\textbf {\bibinfo {volume} {458}},\ \bibinfo {pages} {2634--2647} (\bibinfo {year} {2016})},\ \Eprint {http://arxiv.org/abs/1601.00007} {arXiv:1601.00007 [astro-ph.HE]} \BibitemShut {NoStop}%
\bibitem [{\citenamefont {Ivanova}\ \emph {et~al.}(2008)\citenamefont {Ivanova}, \citenamefont {Heinke}, \citenamefont {Rasio}, \citenamefont {Belczynski},\ and\ \citenamefont {Fregeau}}]{Ivanova:2007bu}%
  \BibitemOpen
  \bibfield  {author} {\bibinfo {author} {\bibfnamefont {N.}~\bibnamefont {Ivanova}}, \bibinfo {author} {\bibfnamefont {C.}~\bibnamefont {Heinke}}, \bibinfo {author} {\bibfnamefont {F.~A.}\ \bibnamefont {Rasio}}, \bibinfo {author} {\bibfnamefont {K.}~\bibnamefont {Belczynski}}, \ and\ \bibinfo {author} {\bibfnamefont {J.}~\bibnamefont {Fregeau}},\ }\bibfield  {title} {\enquote {\bibinfo {title} {{Formation and evolution of compact binaries in globular clusters: II. Binaries with neutron stars}},}\ }\href {\doibase 10.1111/j.1365-2966.2008.13064.x} {\bibfield  {journal} {\bibinfo  {journal} {Mon. Not. Roy. Astron. Soc.}\ }\textbf {\bibinfo {volume} {386}},\ \bibinfo {pages} {553--576} (\bibinfo {year} {2008})},\ \Eprint {http://arxiv.org/abs/0706.4096} {arXiv:0706.4096 [astro-ph]} \BibitemShut {NoStop}%
\bibitem [{\citenamefont {Fragione}\ \emph {et~al.}(2019{\natexlab{a}})\citenamefont {Fragione}, \citenamefont {Grishin}, \citenamefont {Leigh}, \citenamefont {Perets},\ and\ \citenamefont {Perna}}]{Fragione:2018yrb}%
  \BibitemOpen
  \bibfield  {author} {\bibinfo {author} {\bibfnamefont {Giacomo}\ \bibnamefont {Fragione}}, \bibinfo {author} {\bibfnamefont {Evgeni}\ \bibnamefont {Grishin}}, \bibinfo {author} {\bibfnamefont {Nathan W.~C.}\ \bibnamefont {Leigh}}, \bibinfo {author} {\bibfnamefont {Hagai.~B.}\ \bibnamefont {Perets}}, \ and\ \bibinfo {author} {\bibfnamefont {Rosalba}\ \bibnamefont {Perna}},\ }\bibfield  {title} {\enquote {\bibinfo {title} {{Black hole and neutron star mergers in galactic nuclei}},}\ }\href {\doibase 10.1093/mnras/stz1651} {\bibfield  {journal} {\bibinfo  {journal} {Mon. Not. Roy. Astron. Soc.}\ }\textbf {\bibinfo {volume} {488}},\ \bibinfo {pages} {47--63} (\bibinfo {year} {2019}{\natexlab{a}})},\ \Eprint {http://arxiv.org/abs/1811.10627} {arXiv:1811.10627 [astro-ph.GA]} \BibitemShut {NoStop}%
\bibitem [{\citenamefont {Fragione}\ \emph {et~al.}(2019{\natexlab{b}})\citenamefont {Fragione}, \citenamefont {Leigh},\ and\ \citenamefont {Perna}}]{Fragione:2019vgr}%
  \BibitemOpen
  \bibfield  {author} {\bibinfo {author} {\bibfnamefont {Giacomo}\ \bibnamefont {Fragione}}, \bibinfo {author} {\bibfnamefont {Nathan}\ \bibnamefont {Leigh}}, \ and\ \bibinfo {author} {\bibfnamefont {Rosalba}\ \bibnamefont {Perna}},\ }\bibfield  {title} {\enquote {\bibinfo {title} {{Black hole and neutron star mergers in Galactic Nuclei: the role of triples}},}\ }\href {\doibase 10.1093/mnras/stz1803} {\bibfield  {journal} {\bibinfo  {journal} {Mon. Not. Roy. Astron. Soc.}\ }\textbf {\bibinfo {volume} {488}},\ \bibinfo {pages} {2825--2835} (\bibinfo {year} {2019}{\natexlab{b}})},\ \Eprint {http://arxiv.org/abs/1903.09160} {arXiv:1903.09160 [astro-ph.GA]} \BibitemShut {NoStop}%
\bibitem [{\citenamefont {Neumayer}\ \emph {et~al.}(2020)\citenamefont {Neumayer}, \citenamefont {Seth},\ and\ \citenamefont {Boeker}}]{Neumayer:2020gno}%
  \BibitemOpen
  \bibfield  {author} {\bibinfo {author} {\bibfnamefont {Nadine}\ \bibnamefont {Neumayer}}, \bibinfo {author} {\bibfnamefont {Anil}\ \bibnamefont {Seth}}, \ and\ \bibinfo {author} {\bibfnamefont {Torsten}\ \bibnamefont {Boeker}},\ }\bibfield  {title} {\enquote {\bibinfo {title} {{Nuclear star clusters}},}\ }\href {\doibase 10.1007/s00159-020-00125-0} {\bibfield  {journal} {\bibinfo  {journal} {Astron. Astrophys. Rev.}\ }\textbf {\bibinfo {volume} {28}},\ \bibinfo {pages} {4} (\bibinfo {year} {2020})},\ \Eprint {http://arxiv.org/abs/2001.03626} {arXiv:2001.03626 [astro-ph.GA]} \BibitemShut {NoStop}%
\bibitem [{\citenamefont {Santoliquido}\ \emph {et~al.}(2020)\citenamefont {Santoliquido}, \citenamefont {Mapelli}, \citenamefont {Bouffanais}, \citenamefont {Giacobbo}, \citenamefont {Di~Carlo}, \citenamefont {Rastello}, \citenamefont {Artale},\ and\ \citenamefont {Ballone}}]{Santoliquido:2020bry}%
  \BibitemOpen
  \bibfield  {author} {\bibinfo {author} {\bibfnamefont {Filippo}\ \bibnamefont {Santoliquido}}, \bibinfo {author} {\bibfnamefont {Michela}\ \bibnamefont {Mapelli}}, \bibinfo {author} {\bibfnamefont {Yann}\ \bibnamefont {Bouffanais}}, \bibinfo {author} {\bibfnamefont {Nicola}\ \bibnamefont {Giacobbo}}, \bibinfo {author} {\bibfnamefont {Ugo~N.}\ \bibnamefont {Di~Carlo}}, \bibinfo {author} {\bibfnamefont {Sara}\ \bibnamefont {Rastello}}, \bibinfo {author} {\bibfnamefont {M.~Celeste}\ \bibnamefont {Artale}}, \ and\ \bibinfo {author} {\bibfnamefont {Alessandro}\ \bibnamefont {Ballone}},\ }\bibfield  {title} {\enquote {\bibinfo {title} {{The cosmic merger rate density evolution of compact binaries formed in young star clusters and in isolated binaries}},}\ }\href {\doibase 10.3847/1538-4357/ab9b78} {\bibfield  {journal} {\bibinfo  {journal} {Astrophys. J.}\ }\textbf {\bibinfo {volume} {898}},\ \bibinfo {pages} {152} (\bibinfo {year} {2020})},\ \Eprint {http://arxiv.org/abs/2004.09533} {arXiv:2004.09533
  [astro-ph.HE]} \BibitemShut {NoStop}%
\bibitem [{\citenamefont {Rastello}\ \emph {et~al.}(2020)\citenamefont {Rastello}, \citenamefont {Mapelli}, \citenamefont {Di~Carlo}, \citenamefont {Giacobbo}, \citenamefont {Santoliquido}, \citenamefont {Spera}, \citenamefont {Ballone},\ and\ \citenamefont {Iorio}}]{Rastello:2020sru}%
  \BibitemOpen
  \bibfield  {author} {\bibinfo {author} {\bibfnamefont {Sara}\ \bibnamefont {Rastello}}, \bibinfo {author} {\bibfnamefont {Michela}\ \bibnamefont {Mapelli}}, \bibinfo {author} {\bibfnamefont {Ugo~N.}\ \bibnamefont {Di~Carlo}}, \bibinfo {author} {\bibfnamefont {Nicola}\ \bibnamefont {Giacobbo}}, \bibinfo {author} {\bibfnamefont {Filippo}\ \bibnamefont {Santoliquido}}, \bibinfo {author} {\bibfnamefont {Mario}\ \bibnamefont {Spera}}, \bibinfo {author} {\bibfnamefont {Alessandro}\ \bibnamefont {Ballone}}, \ and\ \bibinfo {author} {\bibfnamefont {Giuliano}\ \bibnamefont {Iorio}},\ }\bibfield  {title} {\enquote {\bibinfo {title} {{Dynamics of black hole\textendash{}neutron star binaries in young star clusters}},}\ }\href {\doibase 10.1093/mnras/staa2018} {\bibfield  {journal} {\bibinfo  {journal} {Mon. Not. Roy. Astron. Soc.}\ }\textbf {\bibinfo {volume} {497}},\ \bibinfo {pages} {1563--1570} (\bibinfo {year} {2020})},\ \Eprint {http://arxiv.org/abs/2003.02277} {arXiv:2003.02277 [astro-ph.HE]} \BibitemShut
  {NoStop}%
\bibitem [{\citenamefont {McKernan}\ \emph {et~al.}(2020)\citenamefont {McKernan}, \citenamefont {Ford},\ and\ \citenamefont {O'Shaughnessy}}]{McKernan:2020lgr}%
  \BibitemOpen
  \bibfield  {author} {\bibinfo {author} {\bibfnamefont {B.}~\bibnamefont {McKernan}}, \bibinfo {author} {\bibfnamefont {K.~E.~S.}\ \bibnamefont {Ford}}, \ and\ \bibinfo {author} {\bibfnamefont {R.}~\bibnamefont {O'Shaughnessy}},\ }\bibfield  {title} {\enquote {\bibinfo {title} {{Black hole, neutron star, and white dwarf merger rates in AGN discs}},}\ }\href {\doibase 10.1093/mnras/staa2681} {\bibfield  {journal} {\bibinfo  {journal} {Mon. Not. Roy. Astron. Soc.}\ }\textbf {\bibinfo {volume} {498}},\ \bibinfo {pages} {4088--4094} (\bibinfo {year} {2020})},\ \Eprint {http://arxiv.org/abs/2002.00046} {arXiv:2002.00046 [astro-ph.HE]} \BibitemShut {NoStop}%
\bibitem [{\citenamefont {Mandel}\ and\ \citenamefont {Broekgaarden}(2022)}]{Mandel:2021smh}%
  \BibitemOpen
  \bibfield  {author} {\bibinfo {author} {\bibfnamefont {Ilya}\ \bibnamefont {Mandel}}\ and\ \bibinfo {author} {\bibfnamefont {Floor~S.}\ \bibnamefont {Broekgaarden}},\ }\bibfield  {title} {\enquote {\bibinfo {title} {{Rates of compact object coalescences}},}\ }\href {\doibase 10.1007/s41114-021-00034-3} {\bibfield  {journal} {\bibinfo  {journal} {Living Rev. Rel.}\ }\textbf {\bibinfo {volume} {25}},\ \bibinfo {pages} {1} (\bibinfo {year} {2022})},\ \Eprint {http://arxiv.org/abs/2107.14239} {arXiv:2107.14239 [astro-ph.HE]} \BibitemShut {NoStop}%
\bibitem [{\citenamefont {Zevin}\ \emph {et~al.}(2021{\natexlab{a}})\citenamefont {Zevin}, \citenamefont {Bavera}, \citenamefont {Berry}, \citenamefont {Kalogera}, \citenamefont {Fragos}, \citenamefont {Marchant}, \citenamefont {Rodriguez}, \citenamefont {Antonini}, \citenamefont {Holz},\ and\ \citenamefont {Pankow}}]{Zevin:2020gbd}%
  \BibitemOpen
  \bibfield  {author} {\bibinfo {author} {\bibfnamefont {Michael}\ \bibnamefont {Zevin}}, \bibinfo {author} {\bibfnamefont {Simone~S.}\ \bibnamefont {Bavera}}, \bibinfo {author} {\bibfnamefont {Christopher P.~L.}\ \bibnamefont {Berry}}, \bibinfo {author} {\bibfnamefont {Vicky}\ \bibnamefont {Kalogera}}, \bibinfo {author} {\bibfnamefont {Tassos}\ \bibnamefont {Fragos}}, \bibinfo {author} {\bibfnamefont {Pablo}\ \bibnamefont {Marchant}}, \bibinfo {author} {\bibfnamefont {Carl~L.}\ \bibnamefont {Rodriguez}}, \bibinfo {author} {\bibfnamefont {Fabio}\ \bibnamefont {Antonini}}, \bibinfo {author} {\bibfnamefont {Daniel~E.}\ \bibnamefont {Holz}}, \ and\ \bibinfo {author} {\bibfnamefont {Chris}\ \bibnamefont {Pankow}},\ }\bibfield  {title} {\enquote {\bibinfo {title} {{One Channel to Rule Them All? Constraining the Origins of Binary Black Holes Using Multiple Formation Pathways}},}\ }\href {\doibase 10.3847/1538-4357/abe40e} {\bibfield  {journal} {\bibinfo  {journal} {Astrophys. J.}\ }\textbf {\bibinfo {volume}
  {910}},\ \bibinfo {pages} {152} (\bibinfo {year} {2021}{\natexlab{a}})},\ \Eprint {http://arxiv.org/abs/2011.10057} {arXiv:2011.10057 [astro-ph.HE]} \BibitemShut {NoStop}%
\bibitem [{\citenamefont {Abbott}\ \emph {et~al.}(2023{\natexlab{b}})\citenamefont {Abbott} \emph {et~al.}}]{KAGRA:2021duu}%
  \BibitemOpen
  \bibfield  {author} {\bibinfo {author} {\bibfnamefont {R.}~\bibnamefont {Abbott}} \emph {et~al.} (\bibinfo {collaboration} {KAGRA, VIRGO, LIGO Scientific}),\ }\bibfield  {title} {\enquote {\bibinfo {title} {{Population of Merging Compact Binaries Inferred Using Gravitational Waves through GWTC-3}},}\ }\href {\doibase 10.1103/PhysRevX.13.011048} {\bibfield  {journal} {\bibinfo  {journal} {Phys. Rev. X}\ }\textbf {\bibinfo {volume} {13}},\ \bibinfo {pages} {011048} (\bibinfo {year} {2023}{\natexlab{b}})},\ \Eprint {http://arxiv.org/abs/2111.03634} {arXiv:2111.03634 [astro-ph.HE]} \BibitemShut {NoStop}%
\bibitem [{\citenamefont {Tauris}\ \emph {et~al.}(2017)\citenamefont {Tauris} \emph {et~al.}}]{Tauris:2017omb}%
  \BibitemOpen
  \bibfield  {author} {\bibinfo {author} {\bibfnamefont {T.~M.}\ \bibnamefont {Tauris}} \emph {et~al.},\ }\bibfield  {title} {\enquote {\bibinfo {title} {{Formation of Double Neutron Star Systems}},}\ }\href {\doibase 10.3847/1538-4357/aa7e89} {\bibfield  {journal} {\bibinfo  {journal} {Astrophys. J.}\ }\textbf {\bibinfo {volume} {846}},\ \bibinfo {pages} {170} (\bibinfo {year} {2017})},\ \Eprint {http://arxiv.org/abs/1706.09438} {arXiv:1706.09438 [astro-ph.HE]} \BibitemShut {NoStop}%
\bibitem [{\citenamefont {Bernadich}\ \emph {et~al.}(2023)\citenamefont {Bernadich} \emph {et~al.}}]{Bernadich:2023uru}%
  \BibitemOpen
  \bibfield  {author} {\bibinfo {author} {\bibfnamefont {M.~Colom}\ \bibnamefont {Bernadich}, \bibfnamefont {I}} \emph {et~al.},\ }\bibfield  {title} {\enquote {\bibinfo {title} {{The MPIfR-MeerKAT Galactic Plane Survey - II. The eccentric double neutron star system PSR J1208\ensuremath{-}5936 and a neutron star merger rate update}},}\ }\href {\doibase 10.1051/0004-6361/202346953} {\bibfield  {journal} {\bibinfo  {journal} {Astron. Astrophys.}\ }\textbf {\bibinfo {volume} {678}},\ \bibinfo {pages} {A187} (\bibinfo {year} {2023})},\ \Eprint {http://arxiv.org/abs/2308.16802} {arXiv:2308.16802 [astro-ph.HE]} \BibitemShut {NoStop}%
\bibitem [{\citenamefont {Andrews}\ and\ \citenamefont {Mandel}(2019)}]{Andrews:2019vou}%
  \BibitemOpen
  \bibfield  {author} {\bibinfo {author} {\bibfnamefont {Jeff~J.}\ \bibnamefont {Andrews}}\ and\ \bibinfo {author} {\bibfnamefont {Ilya}\ \bibnamefont {Mandel}},\ }\bibfield  {title} {\enquote {\bibinfo {title} {{Double Neutron Star Populations and Formation Channels}},}\ }\href {\doibase 10.3847/2041-8213/ab2ed1} {\bibfield  {journal} {\bibinfo  {journal} {Astrophys. J. Lett.}\ }\textbf {\bibinfo {volume} {880}},\ \bibinfo {pages} {L8} (\bibinfo {year} {2019})},\ \Eprint {http://arxiv.org/abs/1904.12745} {arXiv:1904.12745 [astro-ph.HE]} \BibitemShut {NoStop}%
\bibitem [{\citenamefont {Zevin}\ \emph {et~al.}(2021{\natexlab{b}})\citenamefont {Zevin}, \citenamefont {Romero-Shaw}, \citenamefont {Kremer}, \citenamefont {Thrane},\ and\ \citenamefont {Lasky}}]{Zevin:2021rtf}%
  \BibitemOpen
  \bibfield  {author} {\bibinfo {author} {\bibfnamefont {Michael}\ \bibnamefont {Zevin}}, \bibinfo {author} {\bibfnamefont {Isobel~M.}\ \bibnamefont {Romero-Shaw}}, \bibinfo {author} {\bibfnamefont {Kyle}\ \bibnamefont {Kremer}}, \bibinfo {author} {\bibfnamefont {Eric}\ \bibnamefont {Thrane}}, \ and\ \bibinfo {author} {\bibfnamefont {Paul~D.}\ \bibnamefont {Lasky}},\ }\bibfield  {title} {\enquote {\bibinfo {title} {{Implications of Eccentric Observations on Binary Black Hole Formation Channels}},}\ }\href {\doibase 10.3847/2041-8213/ac32dc} {\bibfield  {journal} {\bibinfo  {journal} {Astrophys. J. Lett.}\ }\textbf {\bibinfo {volume} {921}},\ \bibinfo {pages} {L43} (\bibinfo {year} {2021}{\natexlab{b}})},\ \Eprint {http://arxiv.org/abs/2106.09042} {arXiv:2106.09042 [astro-ph.HE]} \BibitemShut {NoStop}%
\bibitem [{\citenamefont {Peters}(1964)}]{Peters:1964zz}%
  \BibitemOpen
  \bibfield  {author} {\bibinfo {author} {\bibfnamefont {P.~C.}\ \bibnamefont {Peters}},\ }\bibfield  {title} {\enquote {\bibinfo {title} {{Gravitational Radiation and the Motion of Two Point Masses}},}\ }\href {\doibase 10.1103/PhysRev.136.B1224} {\bibfield  {journal} {\bibinfo  {journal} {Phys. Rev.}\ }\textbf {\bibinfo {volume} {136}},\ \bibinfo {pages} {B1224--B1232} (\bibinfo {year} {1964})}\BibitemShut {NoStop}%
\bibitem [{\citenamefont {Lidov}(1962)}]{LIDOV1962719}%
  \BibitemOpen
  \bibfield  {author} {\bibinfo {author} {\bibfnamefont {M.L.}\ \bibnamefont {Lidov}},\ }\bibfield  {title} {\enquote {\bibinfo {title} {The evolution of orbits of artificial satellites of planets under the action of gravitational perturbations of external bodies},}\ }\href {\doibase https://doi.org/10.1016/0032-0633(62)90129-0} {\bibfield  {journal} {\bibinfo  {journal} {Planetary and Space Science}\ }\textbf {\bibinfo {volume} {9}},\ \bibinfo {pages} {719--759} (\bibinfo {year} {1962})}\BibitemShut {NoStop}%
\bibitem [{\citenamefont {Kozai}(1962)}]{Kozai:1962zz}%
  \BibitemOpen
  \bibfield  {author} {\bibinfo {author} {\bibfnamefont {Yoshihide}\ \bibnamefont {Kozai}},\ }\bibfield  {title} {\enquote {\bibinfo {title} {{Secular perturbations of asteroids with high inclination and eccentricity}},}\ }\href {\doibase 10.1086/108790} {\bibfield  {journal} {\bibinfo  {journal} {Astron. J.}\ }\textbf {\bibinfo {volume} {67}},\ \bibinfo {pages} {591--598} (\bibinfo {year} {1962})}\BibitemShut {NoStop}%
\bibitem [{\citenamefont {Antognini}(2015)}]{Antognini:2015loa}%
  \BibitemOpen
  \bibfield  {author} {\bibinfo {author} {\bibfnamefont {Joseph M.~O.}\ \bibnamefont {Antognini}},\ }\bibfield  {title} {\enquote {\bibinfo {title} {{Timescales of Kozai\textendash{}Lidov oscillations at quadrupole and octupole order in the test particle limit}},}\ }\href {\doibase 10.1093/mnras/stv1552} {\bibfield  {journal} {\bibinfo  {journal} {Mon. Not. Roy. Astron. Soc.}\ }\textbf {\bibinfo {volume} {452}},\ \bibinfo {pages} {3610--3619} (\bibinfo {year} {2015})},\ \Eprint {http://arxiv.org/abs/1504.05957} {arXiv:1504.05957 [astro-ph.EP]} \BibitemShut {NoStop}%
\bibitem [{\citenamefont {Belczynski}\ \emph {et~al.}(2001)\citenamefont {Belczynski}, \citenamefont {Kalogera},\ and\ \citenamefont {Bulik}}]{Belczynski:2001uc}%
  \BibitemOpen
  \bibfield  {author} {\bibinfo {author} {\bibfnamefont {Krzysztof}\ \bibnamefont {Belczynski}}, \bibinfo {author} {\bibfnamefont {Vassiliki}\ \bibnamefont {Kalogera}}, \ and\ \bibinfo {author} {\bibfnamefont {Tomasz}\ \bibnamefont {Bulik}},\ }\bibfield  {title} {\enquote {\bibinfo {title} {{A Comprehensive study of binary compact objects as gravitational wave sources: Evolutionary channels, rates, and physical properties}},}\ }\href {\doibase 10.1086/340304} {\bibfield  {journal} {\bibinfo  {journal} {Astrophys. J.}\ }\textbf {\bibinfo {volume} {572}},\ \bibinfo {pages} {407--431} (\bibinfo {year} {2001})},\ \Eprint {http://arxiv.org/abs/astro-ph/0111452} {arXiv:astro-ph/0111452} \BibitemShut {NoStop}%
\bibitem [{\citenamefont {Belczynski}\ \emph {et~al.}(2018)\citenamefont {Belczynski} \emph {et~al.}}]{Belczynski:2017mqx}%
  \BibitemOpen
  \bibfield  {author} {\bibinfo {author} {\bibfnamefont {K.}~\bibnamefont {Belczynski}} \emph {et~al.},\ }\bibfield  {title} {\enquote {\bibinfo {title} {{The origin of the first neutron star \textendash{} neutron star merger}},}\ }\href {\doibase 10.1051/0004-6361/201732428} {\bibfield  {journal} {\bibinfo  {journal} {Astron. Astrophys.}\ }\textbf {\bibinfo {volume} {615}},\ \bibinfo {pages} {A91} (\bibinfo {year} {2018})},\ \Eprint {http://arxiv.org/abs/1712.00632} {arXiv:1712.00632 [astro-ph.HE]} \BibitemShut {NoStop}%
\bibitem [{\citenamefont {Sedda}(2020)}]{Sedda:2020wzl}%
  \BibitemOpen
  \bibfield  {author} {\bibinfo {author} {\bibfnamefont {Manuel~Arca}\ \bibnamefont {Sedda}},\ }\bibfield  {title} {\enquote {\bibinfo {title} {{Dissecting the properties of neutron star - black hole mergers originating in dense star clusters}},}\ }\href {\doibase 10.1038/s42005-020-0310-x} {\bibfield  {journal} {\bibinfo  {journal} {Commun. Phys.}\ }\textbf {\bibinfo {volume} {3}},\ \bibinfo {pages} {43} (\bibinfo {year} {2020})},\ \Eprint {http://arxiv.org/abs/2003.02279} {arXiv:2003.02279 [astro-ph.GA]} \BibitemShut {NoStop}%
\bibitem [{\citenamefont {Trani}\ \emph {et~al.}(2022)\citenamefont {Trani}, \citenamefont {Rastello}, \citenamefont {Di~Carlo}, \citenamefont {Santoliquido}, \citenamefont {Tanikawa},\ and\ \citenamefont {Mapelli}}]{Trani:2021tan}%
  \BibitemOpen
  \bibfield  {author} {\bibinfo {author} {\bibfnamefont {Alessandro~Alberto}\ \bibnamefont {Trani}}, \bibinfo {author} {\bibfnamefont {Sara}\ \bibnamefont {Rastello}}, \bibinfo {author} {\bibfnamefont {Ugo~N.}\ \bibnamefont {Di~Carlo}}, \bibinfo {author} {\bibfnamefont {Filippo}\ \bibnamefont {Santoliquido}}, \bibinfo {author} {\bibfnamefont {Ataru}\ \bibnamefont {Tanikawa}}, \ and\ \bibinfo {author} {\bibfnamefont {Michela}\ \bibnamefont {Mapelli}},\ }\bibfield  {title} {\enquote {\bibinfo {title} {{Compact object mergers in hierarchical triples from low-mass young star clusters}},}\ }\href {\doibase 10.1093/mnras/stac122} {\bibfield  {journal} {\bibinfo  {journal} {Mon. Not. Roy. Astron. Soc.}\ }\textbf {\bibinfo {volume} {511}},\ \bibinfo {pages} {1362--1372} (\bibinfo {year} {2022})},\ \Eprint {http://arxiv.org/abs/2111.06388} {arXiv:2111.06388 [astro-ph.HE]} \BibitemShut {NoStop}%
\bibitem [{\citenamefont {Hamers}\ and\ \citenamefont {Thompson}(2019)}]{Hamers:2019oeq}%
  \BibitemOpen
  \bibfield  {author} {\bibinfo {author} {\bibfnamefont {Adrian~S.}\ \bibnamefont {Hamers}}\ and\ \bibinfo {author} {\bibfnamefont {Todd~A.}\ \bibnamefont {Thompson}},\ }\bibfield  {title} {\enquote {\bibinfo {title} {{Double neutron star mergers from hierarchical triple-star systems}},}\ }\href {\doibase 10.3847/1538-4357/ab3b06} {\  (\bibinfo {year} {2019}),\ 10.3847/1538-4357/ab3b06},\ \Eprint {http://arxiv.org/abs/1907.08297} {arXiv:1907.08297 [astro-ph.HE]} \BibitemShut {NoStop}%
\bibitem [{\citenamefont {Silsbee}\ and\ \citenamefont {Tremaine}(2017)}]{Silsbee:2016djf}%
  \BibitemOpen
  \bibfield  {author} {\bibinfo {author} {\bibfnamefont {Kedron}\ \bibnamefont {Silsbee}}\ and\ \bibinfo {author} {\bibfnamefont {Scott}\ \bibnamefont {Tremaine}},\ }\bibfield  {title} {\enquote {\bibinfo {title} {{Lidov-Kozai Cycles with Gravitational Radiation: Merging Black Holes in Isolated Triple Systems}},}\ }\href {\doibase 10.3847/1538-4357/aa5729} {\bibfield  {journal} {\bibinfo  {journal} {Astrophys. J.}\ }\textbf {\bibinfo {volume} {836}},\ \bibinfo {pages} {39} (\bibinfo {year} {2017})},\ \Eprint {http://arxiv.org/abs/1608.07642} {arXiv:1608.07642 [astro-ph.HE]} \BibitemShut {NoStop}%
\bibitem [{\citenamefont {Rodriguez}\ and\ \citenamefont {Antonini}(2018)}]{Rodriguez:2018jqu}%
  \BibitemOpen
  \bibfield  {author} {\bibinfo {author} {\bibfnamefont {Carl~L.}\ \bibnamefont {Rodriguez}}\ and\ \bibinfo {author} {\bibfnamefont {Fabio}\ \bibnamefont {Antonini}},\ }\bibfield  {title} {\enquote {\bibinfo {title} {{A Triple Origin for the Heavy and Low-Spin Binary Black Holes Detected by LIGO/Virgo}},}\ }\href {\doibase 10.3847/1538-4357/aacea4} {\bibfield  {journal} {\bibinfo  {journal} {Astrophys. J.}\ }\textbf {\bibinfo {volume} {863}},\ \bibinfo {pages} {7} (\bibinfo {year} {2018})},\ \Eprint {http://arxiv.org/abs/1805.08212} {arXiv:1805.08212 [astro-ph.HE]} \BibitemShut {NoStop}%
\bibitem [{\citenamefont {Kowalska}\ \emph {et~al.}(2011)\citenamefont {Kowalska}, \citenamefont {Bulik}, \citenamefont {Belczynski}, \citenamefont {Dominik},\ and\ \citenamefont {Gondek-Rosinska}}]{Kowalska:2010qg}%
  \BibitemOpen
  \bibfield  {author} {\bibinfo {author} {\bibfnamefont {I.}~\bibnamefont {Kowalska}}, \bibinfo {author} {\bibfnamefont {T.}~\bibnamefont {Bulik}}, \bibinfo {author} {\bibfnamefont {K.}~\bibnamefont {Belczynski}}, \bibinfo {author} {\bibfnamefont {M.}~\bibnamefont {Dominik}}, \ and\ \bibinfo {author} {\bibfnamefont {D.}~\bibnamefont {Gondek-Rosinska}},\ }\bibfield  {title} {\enquote {\bibinfo {title} {{The eccentricity distribution of compact binaries}},}\ }\href {\doibase 10.1051/0004-6361/201015777} {\bibfield  {journal} {\bibinfo  {journal} {Astron. Astrophys.}\ }\textbf {\bibinfo {volume} {527}},\ \bibinfo {pages} {A70} (\bibinfo {year} {2011})},\ \Eprint {http://arxiv.org/abs/1010.0511} {arXiv:1010.0511 [astro-ph.CO]} \BibitemShut {NoStop}%
\bibitem [{\citenamefont {Chaurasia}\ and\ \citenamefont {Bailes}(2005)}]{Chaurasia:2005aq}%
  \BibitemOpen
  \bibfield  {author} {\bibinfo {author} {\bibfnamefont {H.~K.}\ \bibnamefont {Chaurasia}}\ and\ \bibinfo {author} {\bibfnamefont {Matthew}\ \bibnamefont {Bailes}},\ }\bibfield  {title} {\enquote {\bibinfo {title} {{On the eccentricities and merger rates of double neutron star binaries and the creation of double supernovae}},}\ }\href {\doibase 10.1086/444447} {\bibfield  {journal} {\bibinfo  {journal} {Astrophys. J.}\ }\textbf {\bibinfo {volume} {632}},\ \bibinfo {pages} {1054--1059} (\bibinfo {year} {2005})},\ \Eprint {http://arxiv.org/abs/astro-ph/0504021} {arXiv:astro-ph/0504021} \BibitemShut {NoStop}%
\bibitem [{\citenamefont {Richards}\ \emph {et~al.}(2023)\citenamefont {Richards}, \citenamefont {Eldridge}, \citenamefont {Briel}, \citenamefont {Stevance},\ and\ \citenamefont {Willcox}}]{Richards:2022fnq}%
  \BibitemOpen
  \bibfield  {author} {\bibinfo {author} {\bibfnamefont {S.~M.}\ \bibnamefont {Richards}}, \bibinfo {author} {\bibfnamefont {J.~J.}\ \bibnamefont {Eldridge}}, \bibinfo {author} {\bibfnamefont {M.~M.}\ \bibnamefont {Briel}}, \bibinfo {author} {\bibfnamefont {H.~F.}\ \bibnamefont {Stevance}}, \ and\ \bibinfo {author} {\bibfnamefont {R.}~\bibnamefont {Willcox}},\ }\bibfield  {title} {\enquote {\bibinfo {title} {{New constraints on the Bray conservation-of-momentum natal kick model from multiple distinct observations}},}\ }\href {\doibase 10.1093/mnras/stad977} {\bibfield  {journal} {\bibinfo  {journal} {Mon. Not. Roy. Astron. Soc.}\ }\textbf {\bibinfo {volume} {522}},\ \bibinfo {pages} {3972--3985} (\bibinfo {year} {2023})},\ \Eprint {http://arxiv.org/abs/2208.02407} {arXiv:2208.02407 [astro-ph.HE]} \BibitemShut {NoStop}%
\bibitem [{\citenamefont {Abbott}\ \emph {et~al.}(2017)\citenamefont {Abbott} \emph {et~al.}}]{LIGOScientific:2017vwq}%
  \BibitemOpen
  \bibfield  {author} {\bibinfo {author} {\bibfnamefont {B.~P.}\ \bibnamefont {Abbott}} \emph {et~al.} (\bibinfo {collaboration} {LIGO Scientific, Virgo}),\ }\bibfield  {title} {\enquote {\bibinfo {title} {{GW170817: Observation of Gravitational Waves from a Binary Neutron Star Inspiral}},}\ }\href {\doibase 10.1103/PhysRevLett.119.161101} {\bibfield  {journal} {\bibinfo  {journal} {Phys. Rev. Lett.}\ }\textbf {\bibinfo {volume} {119}},\ \bibinfo {pages} {161101} (\bibinfo {year} {2017})},\ \Eprint {http://arxiv.org/abs/1710.05832} {arXiv:1710.05832 [gr-qc]} \BibitemShut {NoStop}%
\bibitem [{\citenamefont {Abbott}\ \emph {et~al.}(2020)\citenamefont {Abbott} \emph {et~al.}}]{LIGOScientific:2020aai}%
  \BibitemOpen
  \bibfield  {author} {\bibinfo {author} {\bibfnamefont {B.~P.}\ \bibnamefont {Abbott}} \emph {et~al.} (\bibinfo {collaboration} {LIGO Scientific, Virgo}),\ }\bibfield  {title} {\enquote {\bibinfo {title} {{GW190425: Observation of a Compact Binary Coalescence with Total Mass $\sim 3.4 M_{\odot}$}},}\ }\href {\doibase 10.3847/2041-8213/ab75f5} {\bibfield  {journal} {\bibinfo  {journal} {Astrophys. J. Lett.}\ }\textbf {\bibinfo {volume} {892}},\ \bibinfo {pages} {L3} (\bibinfo {year} {2020})},\ \Eprint {http://arxiv.org/abs/2001.01761} {arXiv:2001.01761 [astro-ph.HE]} \BibitemShut {NoStop}%
\bibitem [{\citenamefont {Abbott}\ \emph {et~al.}(2021)\citenamefont {Abbott} \emph {et~al.}}]{LIGOScientific:2021qlt}%
  \BibitemOpen
  \bibfield  {author} {\bibinfo {author} {\bibfnamefont {R.}~\bibnamefont {Abbott}} \emph {et~al.} (\bibinfo {collaboration} {LIGO Scientific, KAGRA, VIRGO}),\ }\bibfield  {title} {\enquote {\bibinfo {title} {{Observation of Gravitational Waves from Two Neutron Star\textendash{}Black Hole Coalescences}},}\ }\href {\doibase 10.3847/2041-8213/ac082e} {\bibfield  {journal} {\bibinfo  {journal} {Astrophys. J. Lett.}\ }\textbf {\bibinfo {volume} {915}},\ \bibinfo {pages} {L5} (\bibinfo {year} {2021})},\ \Eprint {http://arxiv.org/abs/2106.15163} {arXiv:2106.15163 [astro-ph.HE]} \BibitemShut {NoStop}%
\bibitem [{\citenamefont {Usman}\ \emph {et~al.}(2016)\citenamefont {Usman} \emph {et~al.}}]{Usman:2015kfa}%
  \BibitemOpen
  \bibfield  {author} {\bibinfo {author} {\bibfnamefont {Samantha~A.}\ \bibnamefont {Usman}} \emph {et~al.},\ }\bibfield  {title} {\enquote {\bibinfo {title} {{The PyCBC search for gravitational waves from compact binary coalescence}},}\ }\href {\doibase 10.1088/0264-9381/33/21/215004} {\bibfield  {journal} {\bibinfo  {journal} {Class. Quant. Grav.}\ }\textbf {\bibinfo {volume} {33}},\ \bibinfo {pages} {215004} (\bibinfo {year} {2016})},\ \Eprint {http://arxiv.org/abs/1508.02357} {arXiv:1508.02357 [gr-qc]} \BibitemShut {NoStop}%
\bibitem [{\citenamefont {Messick}\ \emph {et~al.}(2017)\citenamefont {Messick} \emph {et~al.}}]{Messick:2016aqy}%
  \BibitemOpen
  \bibfield  {author} {\bibinfo {author} {\bibfnamefont {Cody}\ \bibnamefont {Messick}} \emph {et~al.},\ }\bibfield  {title} {\enquote {\bibinfo {title} {{Analysis Framework for the Prompt Discovery of Compact Binary Mergers in Gravitational-wave Data}},}\ }\href {\doibase 10.1103/PhysRevD.95.042001} {\bibfield  {journal} {\bibinfo  {journal} {Phys. Rev. D}\ }\textbf {\bibinfo {volume} {95}},\ \bibinfo {pages} {042001} (\bibinfo {year} {2017})},\ \Eprint {http://arxiv.org/abs/1604.04324} {arXiv:1604.04324 [astro-ph.IM]} \BibitemShut {NoStop}%
\bibitem [{\citenamefont {Aubin}\ \emph {et~al.}(2021)\citenamefont {Aubin} \emph {et~al.}}]{Aubin:2020goo}%
  \BibitemOpen
  \bibfield  {author} {\bibinfo {author} {\bibfnamefont {F.}~\bibnamefont {Aubin}} \emph {et~al.},\ }\bibfield  {title} {\enquote {\bibinfo {title} {{The MBTA pipeline for detecting compact binary coalescences in the third LIGO\textendash{}Virgo observing run}},}\ }\href {\doibase 10.1088/1361-6382/abe913} {\bibfield  {journal} {\bibinfo  {journal} {Class. Quant. Grav.}\ }\textbf {\bibinfo {volume} {38}},\ \bibinfo {pages} {095004} (\bibinfo {year} {2021})},\ \Eprint {http://arxiv.org/abs/2012.11512} {arXiv:2012.11512 [gr-qc]} \BibitemShut {NoStop}%
\bibitem [{\citenamefont {Chu}\ \emph {et~al.}(2022)\citenamefont {Chu} \emph {et~al.}}]{Chu:2020pjv}%
  \BibitemOpen
  \bibfield  {author} {\bibinfo {author} {\bibfnamefont {Qi}~\bibnamefont {Chu}} \emph {et~al.},\ }\bibfield  {title} {\enquote {\bibinfo {title} {{SPIIR online coherent pipeline to search for gravitational waves from compact binary coalescences}},}\ }\href {\doibase 10.1103/PhysRevD.105.024023} {\bibfield  {journal} {\bibinfo  {journal} {Phys. Rev. D}\ }\textbf {\bibinfo {volume} {105}},\ \bibinfo {pages} {024023} (\bibinfo {year} {2022})},\ \Eprint {http://arxiv.org/abs/2011.06787} {arXiv:2011.06787 [gr-qc]} \BibitemShut {NoStop}%
\bibitem [{\citenamefont {Huerta}\ and\ \citenamefont {Brown}(2013)}]{Huerta:2013qb}%
  \BibitemOpen
  \bibfield  {author} {\bibinfo {author} {\bibfnamefont {E.~A.}\ \bibnamefont {Huerta}}\ and\ \bibinfo {author} {\bibfnamefont {Duncan~A.}\ \bibnamefont {Brown}},\ }\bibfield  {title} {\enquote {\bibinfo {title} {{Effect of eccentricity on binary neutron star searches in Advanced LIGO}},}\ }\href {\doibase 10.1103/PhysRevD.87.127501} {\bibfield  {journal} {\bibinfo  {journal} {Phys. Rev. D}\ }\textbf {\bibinfo {volume} {87}},\ \bibinfo {pages} {127501} (\bibinfo {year} {2013})},\ \Eprint {http://arxiv.org/abs/1301.1895} {arXiv:1301.1895 [gr-qc]} \BibitemShut {NoStop}%
\bibitem [{\citenamefont {Lenon}\ \emph {et~al.}(2020)\citenamefont {Lenon}, \citenamefont {Nitz},\ and\ \citenamefont {Brown}}]{Lenon:2020oza}%
  \BibitemOpen
  \bibfield  {author} {\bibinfo {author} {\bibfnamefont {Amber~K.}\ \bibnamefont {Lenon}}, \bibinfo {author} {\bibfnamefont {Alexander~H.}\ \bibnamefont {Nitz}}, \ and\ \bibinfo {author} {\bibfnamefont {Duncan~A.}\ \bibnamefont {Brown}},\ }\bibfield  {title} {\enquote {\bibinfo {title} {{Measuring the eccentricity of GW170817 and GW190425}},}\ }\href {\doibase 10.1093/mnras/staa2120} {\bibfield  {journal} {\bibinfo  {journal} {Mon. Not. Roy. Astron. Soc.}\ }\textbf {\bibinfo {volume} {497}},\ \bibinfo {pages} {1966--1971} (\bibinfo {year} {2020})},\ \Eprint {http://arxiv.org/abs/2005.14146} {arXiv:2005.14146 [astro-ph.HE]} \BibitemShut {NoStop}%
\bibitem [{\citenamefont {Morras}\ \emph {et~al.}(2025)\citenamefont {Morras}, \citenamefont {Pratten},\ and\ \citenamefont {Schmidt}}]{Morras:2025xfu}%
  \BibitemOpen
  \bibfield  {author} {\bibinfo {author} {\bibfnamefont {Gonzalo}\ \bibnamefont {Morras}}, \bibinfo {author} {\bibfnamefont {Geraint}\ \bibnamefont {Pratten}}, \ and\ \bibinfo {author} {\bibfnamefont {Patricia}\ \bibnamefont {Schmidt}},\ }\bibfield  {title} {\enquote {\bibinfo {title} {{Orbital eccentricity in a neutron star - black hole binary}},}\ }\href@noop {} {\  (\bibinfo {year} {2025})},\ \Eprint {http://arxiv.org/abs/2503.15393} {arXiv:2503.15393 [astro-ph.HE]} \BibitemShut {NoStop}%
\bibitem [{\citenamefont {Nitz}\ \emph {et~al.}(2019)\citenamefont {Nitz}, \citenamefont {Lenon},\ and\ \citenamefont {Brown}}]{Nitz:2019spj}%
  \BibitemOpen
  \bibfield  {author} {\bibinfo {author} {\bibfnamefont {Alexander~H.}\ \bibnamefont {Nitz}}, \bibinfo {author} {\bibfnamefont {Amber}\ \bibnamefont {Lenon}}, \ and\ \bibinfo {author} {\bibfnamefont {Duncan~A.}\ \bibnamefont {Brown}},\ }\bibfield  {title} {\enquote {\bibinfo {title} {{Search for Eccentric Binary Neutron Star Mergers in the first and second observing runs of Advanced LIGO}},}\ }\href {\doibase 10.3847/1538-4357/ab6611} {\bibfield  {journal} {\bibinfo  {journal} {Astrophys. J.}\ }\textbf {\bibinfo {volume} {890}},\ \bibinfo {pages} {1} (\bibinfo {year} {2019})},\ \Eprint {http://arxiv.org/abs/1912.05464} {arXiv:1912.05464 [astro-ph.HE]} \BibitemShut {NoStop}%
\bibitem [{\citenamefont {Nitz}\ and\ \citenamefont {Wang}(2021{\natexlab{a}})}]{Nitz:2021vqh}%
  \BibitemOpen
  \bibfield  {author} {\bibinfo {author} {\bibfnamefont {Alexander~H.}\ \bibnamefont {Nitz}}\ and\ \bibinfo {author} {\bibfnamefont {Yi-Fan}\ \bibnamefont {Wang}},\ }\bibfield  {title} {\enquote {\bibinfo {title} {{Search for Gravitational Waves from the Coalescence of Subsolar-Mass Binaries in the First Half of Advanced LIGO and Virgo\textquoteright{}s Third Observing Run}},}\ }\href {\doibase 10.1103/PhysRevLett.127.151101} {\bibfield  {journal} {\bibinfo  {journal} {Phys. Rev. Lett.}\ }\textbf {\bibinfo {volume} {127}},\ \bibinfo {pages} {151101} (\bibinfo {year} {2021}{\natexlab{a}})},\ \Eprint {http://arxiv.org/abs/2106.08979} {arXiv:2106.08979 [astro-ph.HE]} \BibitemShut {NoStop}%
\bibitem [{\citenamefont {Nitz}\ and\ \citenamefont {Wang}(2021{\natexlab{b}})}]{Nitz:2021mzz}%
  \BibitemOpen
  \bibfield  {author} {\bibinfo {author} {\bibfnamefont {Alexander~H.}\ \bibnamefont {Nitz}}\ and\ \bibinfo {author} {\bibfnamefont {Yi-Fan}\ \bibnamefont {Wang}},\ }\bibfield  {title} {\enquote {\bibinfo {title} {{Search for gravitational waves from the coalescence of sub-solar mass and eccentric compact binaries}},}\ }\href {\doibase 10.3847/1538-4357/ac01d9} {\  (\bibinfo {year} {2021}{\natexlab{b}}),\ 10.3847/1538-4357/ac01d9},\ \Eprint {http://arxiv.org/abs/2102.00868} {arXiv:2102.00868 [astro-ph.HE]} \BibitemShut {NoStop}%
\bibitem [{\citenamefont {Abbott}\ \emph {et~al.}(2019{\natexlab{a}})\citenamefont {Abbott} \emph {et~al.}}]{LIGOScientific:2019dag}%
  \BibitemOpen
  \bibfield  {author} {\bibinfo {author} {\bibfnamefont {B.~P.}\ \bibnamefont {Abbott}} \emph {et~al.} (\bibinfo {collaboration} {LIGO Scientific, Virgo}),\ }\bibfield  {title} {\enquote {\bibinfo {title} {{Search for Eccentric Binary Black Hole Mergers with Advanced LIGO and Advanced Virgo during their First and Second Observing Runs}},}\ }\href {\doibase 10.3847/1538-4357/ab3c2d} {\bibfield  {journal} {\bibinfo  {journal} {Astrophys. J.}\ }\textbf {\bibinfo {volume} {883}},\ \bibinfo {pages} {149} (\bibinfo {year} {2019}{\natexlab{a}})},\ \Eprint {http://arxiv.org/abs/1907.09384} {arXiv:1907.09384 [astro-ph.HE]} \BibitemShut {NoStop}%
\bibitem [{\citenamefont {Abac}\ \emph {et~al.}(2023)\citenamefont {Abac} \emph {et~al.}}]{LIGOScientific:2023lpe}%
  \BibitemOpen
  \bibfield  {author} {\bibinfo {author} {\bibfnamefont {A.~G.}\ \bibnamefont {Abac}} \emph {et~al.} (\bibinfo {collaboration} {LIGO Scientific, VIRGO, KAGRA}),\ }\bibfield  {title} {\enquote {\bibinfo {title} {{Search for Eccentric Black Hole Coalescences during the Third Observing Run of LIGO and Virgo}},}\ }\href@noop {} {\  (\bibinfo {year} {2023})},\ \Eprint {http://arxiv.org/abs/2308.03822} {arXiv:2308.03822 [astro-ph.HE]} \BibitemShut {NoStop}%
\bibitem [{\citenamefont {Pal}\ and\ \citenamefont {Nayak}(2024)}]{Pal:2023dyg}%
  \BibitemOpen
  \bibfield  {author} {\bibinfo {author} {\bibfnamefont {Souradeep}\ \bibnamefont {Pal}}\ and\ \bibinfo {author} {\bibfnamefont {K.~Rajesh}\ \bibnamefont {Nayak}},\ }\bibfield  {title} {\enquote {\bibinfo {title} {{Swarm-intelligent search for gravitational waves from eccentric binary mergers}},}\ }\href {\doibase 10.1103/PhysRevD.110.042003} {\bibfield  {journal} {\bibinfo  {journal} {Phys. Rev. D}\ }\textbf {\bibinfo {volume} {110}},\ \bibinfo {pages} {042003} (\bibinfo {year} {2024})},\ \Eprint {http://arxiv.org/abs/2307.03736} {arXiv:2307.03736 [gr-qc]} \BibitemShut {NoStop}%
\bibitem [{\citenamefont {Evans}\ \emph {et~al.}(2021)\citenamefont {Evans} \emph {et~al.}}]{Evans:2021gyd}%
  \BibitemOpen
  \bibfield  {author} {\bibinfo {author} {\bibfnamefont {Matthew}\ \bibnamefont {Evans}} \emph {et~al.},\ }\bibfield  {title} {\enquote {\bibinfo {title} {{A Horizon Study for Cosmic Explorer: Science, Observatories, and Community}},}\ }\href@noop {} {\  (\bibinfo {year} {2021})},\ \Eprint {http://arxiv.org/abs/2109.09882} {arXiv:2109.09882 [astro-ph.IM]} \BibitemShut {NoStop}%
\bibitem [{\citenamefont {Nitz}\ \emph {et~al.}(2017)\citenamefont {Nitz}, \citenamefont {Dent}, \citenamefont {Dal~Canton}, \citenamefont {Fairhurst},\ and\ \citenamefont {Brown}}]{Nitz:2017svb}%
  \BibitemOpen
  \bibfield  {author} {\bibinfo {author} {\bibfnamefont {Alexander~H.}\ \bibnamefont {Nitz}}, \bibinfo {author} {\bibfnamefont {Thomas}\ \bibnamefont {Dent}}, \bibinfo {author} {\bibfnamefont {Tito}\ \bibnamefont {Dal~Canton}}, \bibinfo {author} {\bibfnamefont {Stephen}\ \bibnamefont {Fairhurst}}, \ and\ \bibinfo {author} {\bibfnamefont {Duncan~A.}\ \bibnamefont {Brown}},\ }\bibfield  {title} {\enquote {\bibinfo {title} {{Detecting binary compact-object mergers with gravitational waves: Understanding and Improving the sensitivity of the PyCBC search}},}\ }\href {\doibase 10.3847/1538-4357/aa8f50} {\bibfield  {journal} {\bibinfo  {journal} {Astrophys. J.}\ }\textbf {\bibinfo {volume} {849}},\ \bibinfo {pages} {118} (\bibinfo {year} {2017})},\ \Eprint {http://arxiv.org/abs/1705.01513} {arXiv:1705.01513 [gr-qc]} \BibitemShut {NoStop}%
\bibitem [{\citenamefont {Allen}(2005)}]{Allen:2004gu}%
  \BibitemOpen
  \bibfield  {author} {\bibinfo {author} {\bibfnamefont {Bruce}\ \bibnamefont {Allen}},\ }\bibfield  {title} {\enquote {\bibinfo {title} {{${\chi}^{2}$ time-frequency discriminator for gravitational wave detection}},}\ }\href {\doibase 10.1103/PhysRevD.71.062001} {\bibfield  {journal} {\bibinfo  {journal} {Phys. Rev. D}\ }\textbf {\bibinfo {volume} {71}},\ \bibinfo {pages} {062001} (\bibinfo {year} {2005})},\ \Eprint {http://arxiv.org/abs/gr-qc/0405045} {arXiv:gr-qc/0405045} \BibitemShut {NoStop}%
\bibitem [{\citenamefont {Nitz}(2018)}]{Nitz:2017lco}%
  \BibitemOpen
  \bibfield  {author} {\bibinfo {author} {\bibfnamefont {Alexander~Harvey}\ \bibnamefont {Nitz}},\ }\bibfield  {title} {\enquote {\bibinfo {title} {{Distinguishing short duration noise transients in LIGO data to improve the PyCBC search for gravitational waves from high mass binary black hole mergers}},}\ }\href {\doibase 10.1088/1361-6382/aaa13d} {\bibfield  {journal} {\bibinfo  {journal} {Class. Quant. Grav.}\ }\textbf {\bibinfo {volume} {35}},\ \bibinfo {pages} {035016} (\bibinfo {year} {2018})},\ \Eprint {http://arxiv.org/abs/1709.08974} {arXiv:1709.08974 [gr-qc]} \BibitemShut {NoStop}%
\bibitem [{\citenamefont {Davies}\ \emph {et~al.}(2020)\citenamefont {Davies}, \citenamefont {Dent}, \citenamefont {T\'apai}, \citenamefont {Harry}, \citenamefont {McIsaac},\ and\ \citenamefont {Nitz}}]{Davies:2020tsx}%
  \BibitemOpen
  \bibfield  {author} {\bibinfo {author} {\bibfnamefont {Gareth~S.}\ \bibnamefont {Davies}}, \bibinfo {author} {\bibfnamefont {Thomas}\ \bibnamefont {Dent}}, \bibinfo {author} {\bibfnamefont {M\'arton}\ \bibnamefont {T\'apai}}, \bibinfo {author} {\bibfnamefont {Ian}\ \bibnamefont {Harry}}, \bibinfo {author} {\bibfnamefont {Connor}\ \bibnamefont {McIsaac}}, \ and\ \bibinfo {author} {\bibfnamefont {Alexander~H.}\ \bibnamefont {Nitz}},\ }\bibfield  {title} {\enquote {\bibinfo {title} {{Extending the PyCBC search for gravitational waves from compact binary mergers to a global network}},}\ }\href {\doibase 10.1103/PhysRevD.102.022004} {\bibfield  {journal} {\bibinfo  {journal} {Phys. Rev. D}\ }\textbf {\bibinfo {volume} {102}},\ \bibinfo {pages} {022004} (\bibinfo {year} {2020})},\ \Eprint {http://arxiv.org/abs/2002.08291} {arXiv:2002.08291 [astro-ph.HE]} \BibitemShut {NoStop}%
\bibitem [{\citenamefont {Was}\ \emph {et~al.}(2010)\citenamefont {Was}, \citenamefont {Bizouard}, \citenamefont {Brisson}, \citenamefont {Cavalier}, \citenamefont {Davier}, \citenamefont {Hello}, \citenamefont {Leroy}, \citenamefont {Robinet},\ and\ \citenamefont {Vavoulidis}}]{Was:2009vh}%
  \BibitemOpen
  \bibfield  {author} {\bibinfo {author} {\bibfnamefont {Michal}\ \bibnamefont {Was}}, \bibinfo {author} {\bibfnamefont {Marie-Anne}\ \bibnamefont {Bizouard}}, \bibinfo {author} {\bibfnamefont {Violette}\ \bibnamefont {Brisson}}, \bibinfo {author} {\bibfnamefont {Fabien}\ \bibnamefont {Cavalier}}, \bibinfo {author} {\bibfnamefont {Michel}\ \bibnamefont {Davier}}, \bibinfo {author} {\bibfnamefont {Patrice}\ \bibnamefont {Hello}}, \bibinfo {author} {\bibfnamefont {Nicolas}\ \bibnamefont {Leroy}}, \bibinfo {author} {\bibfnamefont {Florent}\ \bibnamefont {Robinet}}, \ and\ \bibinfo {author} {\bibfnamefont {Miltiadis}\ \bibnamefont {Vavoulidis}},\ }\bibfield  {title} {\enquote {\bibinfo {title} {{On the background estimation by time slides in a network of gravitational wave detectors}},}\ }\href {\doibase 10.1088/0264-9381/27/1/015005} {\bibfield  {journal} {\bibinfo  {journal} {Class. Quant. Grav.}\ }\textbf {\bibinfo {volume} {27}},\ \bibinfo {pages} {015005} (\bibinfo {year} {2010})},\ \Eprint
  {http://arxiv.org/abs/0906.2120} {arXiv:0906.2120 [gr-qc]} \BibitemShut {NoStop}%
\bibitem [{\citenamefont {Harry}\ \emph {et~al.}(2009)\citenamefont {Harry}, \citenamefont {Allen},\ and\ \citenamefont {Sathyaprakash}}]{Harry:2009ea}%
  \BibitemOpen
  \bibfield  {author} {\bibinfo {author} {\bibfnamefont {Ian~W.}\ \bibnamefont {Harry}}, \bibinfo {author} {\bibfnamefont {Bruce}\ \bibnamefont {Allen}}, \ and\ \bibinfo {author} {\bibfnamefont {B.~S.}\ \bibnamefont {Sathyaprakash}},\ }\bibfield  {title} {\enquote {\bibinfo {title} {{A Stochastic template placement algorithm for gravitational wave data analysis}},}\ }\href {\doibase 10.1103/PhysRevD.80.104014} {\bibfield  {journal} {\bibinfo  {journal} {Phys. Rev. D}\ }\textbf {\bibinfo {volume} {80}},\ \bibinfo {pages} {104014} (\bibinfo {year} {2009})},\ \Eprint {http://arxiv.org/abs/0908.2090} {arXiv:0908.2090 [gr-qc]} \BibitemShut {NoStop}%
\bibitem [{\citenamefont {Babak}(2008)}]{Babak:2008rb}%
  \BibitemOpen
  \bibfield  {author} {\bibinfo {author} {\bibfnamefont {Stanislav}\ \bibnamefont {Babak}},\ }\bibfield  {title} {\enquote {\bibinfo {title} {{Building a stochastic template bank for detecting massive black hole binaries}},}\ }\href {\doibase 10.1088/0264-9381/25/19/195011} {\bibfield  {journal} {\bibinfo  {journal} {Class. Quant. Grav.}\ }\textbf {\bibinfo {volume} {25}},\ \bibinfo {pages} {195011} (\bibinfo {year} {2008})},\ \Eprint {http://arxiv.org/abs/0801.4070} {arXiv:0801.4070 [gr-qc]} \BibitemShut {NoStop}%
\bibitem [{\citenamefont {Moore}\ \emph {et~al.}(2016)\citenamefont {Moore}, \citenamefont {Favata}, \citenamefont {Arun},\ and\ \citenamefont {Mishra}}]{Moore:2016qxz}%
  \BibitemOpen
  \bibfield  {author} {\bibinfo {author} {\bibfnamefont {Blake}\ \bibnamefont {Moore}}, \bibinfo {author} {\bibfnamefont {Marc}\ \bibnamefont {Favata}}, \bibinfo {author} {\bibfnamefont {K.~G.}\ \bibnamefont {Arun}}, \ and\ \bibinfo {author} {\bibfnamefont {Chandra~Kant}\ \bibnamefont {Mishra}},\ }\bibfield  {title} {\enquote {\bibinfo {title} {{Gravitational-wave phasing for low-eccentricity inspiralling compact binaries to 3PN order}},}\ }\href {\doibase 10.1103/PhysRevD.93.124061} {\bibfield  {journal} {\bibinfo  {journal} {Phys. Rev. D}\ }\textbf {\bibinfo {volume} {93}},\ \bibinfo {pages} {124061} (\bibinfo {year} {2016})},\ \Eprint {http://arxiv.org/abs/1605.00304} {arXiv:1605.00304 [gr-qc]} \BibitemShut {NoStop}%
\bibitem [{\citenamefont {{LIGO Scientific Collaboration}}\ \emph {et~al.}(2018)\citenamefont {{LIGO Scientific Collaboration}}, \citenamefont {{Virgo Collaboration}},\ and\ \citenamefont {{KAGRA Collaboration}}}]{lalsuite}%
  \BibitemOpen
  \bibfield  {author} {\bibinfo {author} {\bibnamefont {{LIGO Scientific Collaboration}}}, \bibinfo {author} {\bibnamefont {{Virgo Collaboration}}}, \ and\ \bibinfo {author} {\bibnamefont {{KAGRA Collaboration}}},\ }\href {\doibase 10.7935/GT1W-FZ16} {\enquote {\bibinfo {title} {{LVK} {A}lgorithm {L}ibrary - {LALS}uite},}\ }\bibinfo {howpublished} {Free software (GPL)} (\bibinfo {year} {2018})\BibitemShut {NoStop}%
\bibitem [{\citenamefont {Boh\'e}\ \emph {et~al.}(2013)\citenamefont {Boh\'e}, \citenamefont {Marsat},\ and\ \citenamefont {Blanchet}}]{Bohe:2013cla}%
  \BibitemOpen
  \bibfield  {author} {\bibinfo {author} {\bibfnamefont {Alejandro}\ \bibnamefont {Boh\'e}}, \bibinfo {author} {\bibfnamefont {Sylvain}\ \bibnamefont {Marsat}}, \ and\ \bibinfo {author} {\bibfnamefont {Luc}\ \bibnamefont {Blanchet}},\ }\bibfield  {title} {\enquote {\bibinfo {title} {{Next-to-next-to-leading order spin\textendash{}orbit effects in the gravitational wave flux and orbital phasing of compact binaries}},}\ }\href {\doibase 10.1088/0264-9381/30/13/135009} {\bibfield  {journal} {\bibinfo  {journal} {Class. Quant. Grav.}\ }\textbf {\bibinfo {volume} {30}},\ \bibinfo {pages} {135009} (\bibinfo {year} {2013})},\ \Eprint {http://arxiv.org/abs/1303.7412} {arXiv:1303.7412 [gr-qc]} \BibitemShut {NoStop}%
\bibitem [{\citenamefont {Boh\'e}\ \emph {et~al.}(2015)\citenamefont {Boh\'e}, \citenamefont {Faye}, \citenamefont {Marsat},\ and\ \citenamefont {Porter}}]{Bohe:2015ana}%
  \BibitemOpen
  \bibfield  {author} {\bibinfo {author} {\bibfnamefont {Alejandro}\ \bibnamefont {Boh\'e}}, \bibinfo {author} {\bibfnamefont {Guillaume}\ \bibnamefont {Faye}}, \bibinfo {author} {\bibfnamefont {Sylvain}\ \bibnamefont {Marsat}}, \ and\ \bibinfo {author} {\bibfnamefont {Edward~K.}\ \bibnamefont {Porter}},\ }\bibfield  {title} {\enquote {\bibinfo {title} {{Quadratic-in-spin effects in the orbital dynamics and gravitational-wave energy flux of compact binaries at the 3PN order}},}\ }\href {\doibase 10.1088/0264-9381/32/19/195010} {\bibfield  {journal} {\bibinfo  {journal} {Class. Quant. Grav.}\ }\textbf {\bibinfo {volume} {32}},\ \bibinfo {pages} {195010} (\bibinfo {year} {2015})},\ \Eprint {http://arxiv.org/abs/1501.01529} {arXiv:1501.01529 [gr-qc]} \BibitemShut {NoStop}%
\bibitem [{\citenamefont {Abbott}\ \emph {et~al.}(2023{\natexlab{c}})\citenamefont {Abbott} \emph {et~al.}}]{KAGRA:2023pio}%
  \BibitemOpen
  \bibfield  {author} {\bibinfo {author} {\bibfnamefont {R.}~\bibnamefont {Abbott}} \emph {et~al.} (\bibinfo {collaboration} {KAGRA, VIRGO, LIGO Scientific}),\ }\bibfield  {title} {\enquote {\bibinfo {title} {{Open Data from the Third Observing Run of LIGO, Virgo, KAGRA, and GEO}},}\ }\href {\doibase 10.3847/1538-4365/acdc9f} {\bibfield  {journal} {\bibinfo  {journal} {Astrophys. J. Suppl.}\ }\textbf {\bibinfo {volume} {267}},\ \bibinfo {pages} {29} (\bibinfo {year} {2023}{\natexlab{c}})},\ \Eprint {http://arxiv.org/abs/2302.03676} {arXiv:2302.03676 [gr-qc]} \BibitemShut {NoStop}%
\bibitem [{\citenamefont {Dhurkunde}(2023)}]{github}%
  \BibitemOpen
  \bibfield  {author} {\bibinfo {author} {\bibfnamefont {Rahul}\ \bibnamefont {Dhurkunde}},\ }\href@noop {} {\enquote {\bibinfo {title} {{Data Release: Eccentric search for Neutron star-black hole (NSBH) and binary neutron star (BNS) mergers within O3 advanced LIGO and advanced VIRGO data}},}\ }\bibinfo {howpublished} {\url{https://github.com/rahuldhurkunde/Eccentric-search-O3}} (\bibinfo {year} {2023})\BibitemShut {NoStop}%
\bibitem [{\citenamefont {Biswas}\ \emph {et~al.}(2009)\citenamefont {Biswas}, \citenamefont {Brady}, \citenamefont {Creighton},\ and\ \citenamefont {Fairhurst}}]{Biswas:2007ni}%
  \BibitemOpen
  \bibfield  {author} {\bibinfo {author} {\bibfnamefont {Rahul}\ \bibnamefont {Biswas}}, \bibinfo {author} {\bibfnamefont {Patrick~R.}\ \bibnamefont {Brady}}, \bibinfo {author} {\bibfnamefont {Jolien D.~E.}\ \bibnamefont {Creighton}}, \ and\ \bibinfo {author} {\bibfnamefont {Stephen}\ \bibnamefont {Fairhurst}},\ }\bibfield  {title} {\enquote {\bibinfo {title} {{The Loudest event statistic: General formulation, properties and applications}},}\ }\href {\doibase 10.1088/0264-9381/26/17/175009} {\bibfield  {journal} {\bibinfo  {journal} {Class. Quant. Grav.}\ }\textbf {\bibinfo {volume} {26}},\ \bibinfo {pages} {175009} (\bibinfo {year} {2009})},\ \bibinfo {note} {[Erratum: Class.Quant.Grav. 30, 079502 (2013)]},\ \Eprint {http://arxiv.org/abs/0710.0465} {arXiv:0710.0465 [gr-qc]} \BibitemShut {NoStop}%
\bibitem [{\citenamefont {Tiwari}(2018)}]{Tiwari:2017ndi}%
  \BibitemOpen
  \bibfield  {author} {\bibinfo {author} {\bibfnamefont {Vaibhav}\ \bibnamefont {Tiwari}},\ }\bibfield  {title} {\enquote {\bibinfo {title} {{Estimation of the Sensitive Volume for Gravitational-wave Source Populations Using Weighted Monte Carlo Integration}},}\ }\href {\doibase 10.1088/1361-6382/aac89d} {\bibfield  {journal} {\bibinfo  {journal} {Class. Quant. Grav.}\ }\textbf {\bibinfo {volume} {35}},\ \bibinfo {pages} {145009} (\bibinfo {year} {2018})},\ \Eprint {http://arxiv.org/abs/1712.00482} {arXiv:1712.00482 [astro-ph.HE]} \BibitemShut {NoStop}%
\bibitem [{\citenamefont {Madau}\ and\ \citenamefont {Fragos}(2017)}]{Madau:2016jbv}%
  \BibitemOpen
  \bibfield  {author} {\bibinfo {author} {\bibfnamefont {Piero}\ \bibnamefont {Madau}}\ and\ \bibinfo {author} {\bibfnamefont {Tassos}\ \bibnamefont {Fragos}},\ }\bibfield  {title} {\enquote {\bibinfo {title} {{Radiation Backgrounds at Cosmic Dawn: X-Rays from Compact Binaries}},}\ }\href {\doibase 10.3847/1538-4357/aa6af9} {\bibfield  {journal} {\bibinfo  {journal} {Astrophys. J.}\ }\textbf {\bibinfo {volume} {840}},\ \bibinfo {pages} {39} (\bibinfo {year} {2017})},\ \Eprint {http://arxiv.org/abs/1606.07887} {arXiv:1606.07887 [astro-ph.GA]} \BibitemShut {NoStop}%
\bibitem [{\citenamefont {Zhu}\ \emph {et~al.}(2021)\citenamefont {Zhu} \emph {et~al.}}]{Zhu:2020ffa}%
  \BibitemOpen
  \bibfield  {author} {\bibinfo {author} {\bibfnamefont {Jin-Ping}\ \bibnamefont {Zhu}} \emph {et~al.},\ }\bibfield  {title} {\enquote {\bibinfo {title} {{Kilonova Emission from Black Hole\textendash{}Neutron Star Mergers. II. Luminosity Function and Implications for Target-of-opportunity Observations of Gravitational-wave Triggers and Blind Searches}},}\ }\href {\doibase 10.3847/1538-4357/abfe5e} {\bibfield  {journal} {\bibinfo  {journal} {Astrophys. J.}\ }\textbf {\bibinfo {volume} {917}},\ \bibinfo {pages} {24} (\bibinfo {year} {2021})},\ \Eprint {http://arxiv.org/abs/2011.02717} {arXiv:2011.02717 [astro-ph.HE]} \BibitemShut {NoStop}%
\bibitem [{\citenamefont {Wu}\ and\ \citenamefont {Nitz}(2023)}]{Wu:2022pyg}%
  \BibitemOpen
  \bibfield  {author} {\bibinfo {author} {\bibfnamefont {Shichao}\ \bibnamefont {Wu}}\ and\ \bibinfo {author} {\bibfnamefont {Alexander~H.}\ \bibnamefont {Nitz}},\ }\bibfield  {title} {\enquote {\bibinfo {title} {{Mock data study for next-generation ground-based detectors: The performance loss of matched filtering due to correlated confusion noise}},}\ }\href {\doibase 10.1103/PhysRevD.107.063022} {\bibfield  {journal} {\bibinfo  {journal} {Phys. Rev. D}\ }\textbf {\bibinfo {volume} {107}},\ \bibinfo {pages} {063022} (\bibinfo {year} {2023})},\ \Eprint {http://arxiv.org/abs/2209.03135} {arXiv:2209.03135 [astro-ph.IM]} \BibitemShut {NoStop}%
\bibitem [{\citenamefont {Abbott}\ \emph {et~al.}(2019{\natexlab{b}})\citenamefont {Abbott} \emph {et~al.}}]{LIGOScientific:2018hze}%
  \BibitemOpen
  \bibfield  {author} {\bibinfo {author} {\bibfnamefont {B.~P.}\ \bibnamefont {Abbott}} \emph {et~al.} (\bibinfo {collaboration} {LIGO Scientific, Virgo}),\ }\bibfield  {title} {\enquote {\bibinfo {title} {{Properties of the binary neutron star merger GW170817}},}\ }\href {\doibase 10.1103/PhysRevX.9.011001} {\bibfield  {journal} {\bibinfo  {journal} {Phys. Rev. X}\ }\textbf {\bibinfo {volume} {9}},\ \bibinfo {pages} {011001} (\bibinfo {year} {2019}{\natexlab{b}})},\ \Eprint {http://arxiv.org/abs/1805.11579} {arXiv:1805.11579 [gr-qc]} \BibitemShut {NoStop}%
\bibitem [{\citenamefont {Pratten}\ \emph {et~al.}(2021)\citenamefont {Pratten} \emph {et~al.}}]{Pratten:2020ceb}%
  \BibitemOpen
  \bibfield  {author} {\bibinfo {author} {\bibfnamefont {Geraint}\ \bibnamefont {Pratten}} \emph {et~al.},\ }\bibfield  {title} {\enquote {\bibinfo {title} {{Computationally efficient models for the dominant and subdominant harmonic modes of precessing binary black holes}},}\ }\href {\doibase 10.1103/PhysRevD.103.104056} {\bibfield  {journal} {\bibinfo  {journal} {Phys. Rev. D}\ }\textbf {\bibinfo {volume} {103}},\ \bibinfo {pages} {104056} (\bibinfo {year} {2021})},\ \Eprint {http://arxiv.org/abs/2004.06503} {arXiv:2004.06503 [gr-qc]} \BibitemShut {NoStop}%
\bibitem [{\citenamefont {Pratten}\ \emph {et~al.}(2020)\citenamefont {Pratten}, \citenamefont {Husa}, \citenamefont {Garcia-Quiros}, \citenamefont {Colleoni}, \citenamefont {Ramos-Buades}, \citenamefont {Estelles},\ and\ \citenamefont {Jaume}}]{Pratten:2020fqn}%
  \BibitemOpen
  \bibfield  {author} {\bibinfo {author} {\bibfnamefont {Geraint}\ \bibnamefont {Pratten}}, \bibinfo {author} {\bibfnamefont {Sascha}\ \bibnamefont {Husa}}, \bibinfo {author} {\bibfnamefont {Cecilio}\ \bibnamefont {Garcia-Quiros}}, \bibinfo {author} {\bibfnamefont {Marta}\ \bibnamefont {Colleoni}}, \bibinfo {author} {\bibfnamefont {Antoni}\ \bibnamefont {Ramos-Buades}}, \bibinfo {author} {\bibfnamefont {Hector}\ \bibnamefont {Estelles}}, \ and\ \bibinfo {author} {\bibfnamefont {Rafel}\ \bibnamefont {Jaume}},\ }\bibfield  {title} {\enquote {\bibinfo {title} {{Setting the cornerstone for a family of models for gravitational waves from compact binaries: The dominant harmonic for nonprecessing quasicircular black holes}},}\ }\href {\doibase 10.1103/PhysRevD.102.064001} {\bibfield  {journal} {\bibinfo  {journal} {Phys. Rev. D}\ }\textbf {\bibinfo {volume} {102}},\ \bibinfo {pages} {064001} (\bibinfo {year} {2020})},\ \Eprint {http://arxiv.org/abs/2001.11412} {arXiv:2001.11412 [gr-qc]} \BibitemShut {NoStop}%
\bibitem [{\citenamefont {Cullen}\ \emph {et~al.}(2017)\citenamefont {Cullen}, \citenamefont {Harry}, \citenamefont {Read},\ and\ \citenamefont {Flynn}}]{Cullen:2017oaz}%
  \BibitemOpen
  \bibfield  {author} {\bibinfo {author} {\bibfnamefont {Torrey}\ \bibnamefont {Cullen}}, \bibinfo {author} {\bibfnamefont {Ian}\ \bibnamefont {Harry}}, \bibinfo {author} {\bibfnamefont {Jocelyn}\ \bibnamefont {Read}}, \ and\ \bibinfo {author} {\bibfnamefont {Eric}\ \bibnamefont {Flynn}},\ }\bibfield  {title} {\enquote {\bibinfo {title} {{Matter Effects on LIGO/Virgo Searches for Gravitational Waves from Merging Neutron Stars}},}\ }\href {\doibase 10.1088/1361-6382/aa9424} {\bibfield  {journal} {\bibinfo  {journal} {Class. Quant. Grav.}\ }\textbf {\bibinfo {volume} {34}},\ \bibinfo {pages} {245003} (\bibinfo {year} {2017})},\ \Eprint {http://arxiv.org/abs/1708.04359} {arXiv:1708.04359 [gr-qc]} \BibitemShut {NoStop}%
\bibitem [{\citenamefont {Nagar}\ and\ \citenamefont {Rettegno}(2021)}]{Nagar:2021xnh}%
  \BibitemOpen
  \bibfield  {author} {\bibinfo {author} {\bibfnamefont {Alessandro}\ \bibnamefont {Nagar}}\ and\ \bibinfo {author} {\bibfnamefont {Piero}\ \bibnamefont {Rettegno}},\ }\bibfield  {title} {\enquote {\bibinfo {title} {{Next generation: Impact of high-order analytical information on effective one body waveform models for noncircularized, spin-aligned black hole binaries}},}\ }\href {\doibase 10.1103/PhysRevD.104.104004} {\bibfield  {journal} {\bibinfo  {journal} {Phys. Rev. D}\ }\textbf {\bibinfo {volume} {104}},\ \bibinfo {pages} {104004} (\bibinfo {year} {2021})},\ \Eprint {http://arxiv.org/abs/2108.02043} {arXiv:2108.02043 [gr-qc]} \BibitemShut {NoStop}%
\bibitem [{\citenamefont {Srivastava}\ \emph {et~al.}(2022)\citenamefont {Srivastava}, \citenamefont {Davis}, \citenamefont {Kuns}, \citenamefont {Landry}, \citenamefont {Ballmer}, \citenamefont {Evans}, \citenamefont {Hall}, \citenamefont {Read},\ and\ \citenamefont {Sathyaprakash}}]{Srivastava:2022slt}%
  \BibitemOpen
  \bibfield  {author} {\bibinfo {author} {\bibfnamefont {Varun}\ \bibnamefont {Srivastava}}, \bibinfo {author} {\bibfnamefont {Derek}\ \bibnamefont {Davis}}, \bibinfo {author} {\bibfnamefont {Kevin}\ \bibnamefont {Kuns}}, \bibinfo {author} {\bibfnamefont {Philippe}\ \bibnamefont {Landry}}, \bibinfo {author} {\bibfnamefont {Stefan}\ \bibnamefont {Ballmer}}, \bibinfo {author} {\bibfnamefont {Matthew}\ \bibnamefont {Evans}}, \bibinfo {author} {\bibfnamefont {Evan~D.}\ \bibnamefont {Hall}}, \bibinfo {author} {\bibfnamefont {Jocelyn}\ \bibnamefont {Read}}, \ and\ \bibinfo {author} {\bibfnamefont {B.~S.}\ \bibnamefont {Sathyaprakash}},\ }\bibfield  {title} {\enquote {\bibinfo {title} {{Science-driven Tunable Design of Cosmic Explorer Detectors}},}\ }\href {\doibase 10.3847/1538-4357/ac5f04} {\bibfield  {journal} {\bibinfo  {journal} {Astrophys. J.}\ }\textbf {\bibinfo {volume} {931}},\ \bibinfo {pages} {22} (\bibinfo {year} {2022})},\ \Eprint {http://arxiv.org/abs/2201.10668} {arXiv:2201.10668 [gr-qc]} \BibitemShut
  {NoStop}%
\bibitem [{\citenamefont {LIGO~Scientific}(2024)}]{post-o5-curves}%
  \BibitemOpen
  \bibfield  {author} {\bibinfo {author} {\bibfnamefont {Virgo}\ \bibnamefont {LIGO~Scientific}},\ }\href@noop {} {\enquote {\bibinfo {title} {{Report from the LSC Post-O5 Study Group}},}\ }\bibinfo {howpublished} {\url{https://dcc.ligo.org/public/0183/T2200287/003/T2200287v3_PO5report.pdf}} (\bibinfo {year} {2024}),\ \bibinfo {note} {[Accessed 30-08-2024]}\BibitemShut {NoStop}%
\bibitem [{\citenamefont {Husa}\ \emph {et~al.}(2016)\citenamefont {Husa}, \citenamefont {Khan}, \citenamefont {Hannam}, \citenamefont {P\"urrer}, \citenamefont {Ohme}, \citenamefont {Jim\'enez~Forteza},\ and\ \citenamefont {Boh\'e}}]{Husa:2015iqa}%
  \BibitemOpen
  \bibfield  {author} {\bibinfo {author} {\bibfnamefont {Sascha}\ \bibnamefont {Husa}}, \bibinfo {author} {\bibfnamefont {Sebastian}\ \bibnamefont {Khan}}, \bibinfo {author} {\bibfnamefont {Mark}\ \bibnamefont {Hannam}}, \bibinfo {author} {\bibfnamefont {Michael}\ \bibnamefont {P\"urrer}}, \bibinfo {author} {\bibfnamefont {Frank}\ \bibnamefont {Ohme}}, \bibinfo {author} {\bibfnamefont {Xisco}\ \bibnamefont {Jim\'enez~Forteza}}, \ and\ \bibinfo {author} {\bibfnamefont {Alejandro}\ \bibnamefont {Boh\'e}},\ }\bibfield  {title} {\enquote {\bibinfo {title} {{Frequency-domain gravitational waves from nonprecessing black-hole binaries. I. New numerical waveforms and anatomy of the signal}},}\ }\href {\doibase 10.1103/PhysRevD.93.044006} {\bibfield  {journal} {\bibinfo  {journal} {Phys. Rev. D}\ }\textbf {\bibinfo {volume} {93}},\ \bibinfo {pages} {044006} (\bibinfo {year} {2016})},\ \Eprint {http://arxiv.org/abs/1508.07250} {arXiv:1508.07250 [gr-qc]} \BibitemShut {NoStop}%
\bibitem [{\citenamefont {Khan}\ \emph {et~al.}(2016)\citenamefont {Khan}, \citenamefont {Husa}, \citenamefont {Hannam}, \citenamefont {Ohme}, \citenamefont {P\"urrer}, \citenamefont {Jim\'enez~Forteza},\ and\ \citenamefont {Boh\'e}}]{Khan:2015jqa}%
  \BibitemOpen
  \bibfield  {author} {\bibinfo {author} {\bibfnamefont {Sebastian}\ \bibnamefont {Khan}}, \bibinfo {author} {\bibfnamefont {Sascha}\ \bibnamefont {Husa}}, \bibinfo {author} {\bibfnamefont {Mark}\ \bibnamefont {Hannam}}, \bibinfo {author} {\bibfnamefont {Frank}\ \bibnamefont {Ohme}}, \bibinfo {author} {\bibfnamefont {Michael}\ \bibnamefont {P\"urrer}}, \bibinfo {author} {\bibfnamefont {Xisco}\ \bibnamefont {Jim\'enez~Forteza}}, \ and\ \bibinfo {author} {\bibfnamefont {Alejandro}\ \bibnamefont {Boh\'e}},\ }\bibfield  {title} {\enquote {\bibinfo {title} {{Frequency-domain gravitational waves from nonprecessing black-hole binaries. II. A phenomenological model for the advanced detector era}},}\ }\href {\doibase 10.1103/PhysRevD.93.044007} {\bibfield  {journal} {\bibinfo  {journal} {Phys. Rev. D}\ }\textbf {\bibinfo {volume} {93}},\ \bibinfo {pages} {044007} (\bibinfo {year} {2016})},\ \Eprint {http://arxiv.org/abs/1508.07253} {arXiv:1508.07253 [gr-qc]} \BibitemShut {NoStop}%
\bibitem [{Ash(2023)}]{Asharp_sensitivity}%
  \BibitemOpen
  \href@noop {} {\enquote {\bibinfo {title} {Asharp strain sensitivity},}\ }\bibinfo {howpublished} {\url{https://dcc.ligo.org/LIGO-T2300041/public}} (\bibinfo {year} {2023}),\ \bibinfo {note} {accessed: 27-01-2023}\BibitemShut {NoStop}%
\bibitem [{CE_(2023)}]{CE_sensitivity}%
  \BibitemOpen
  \href@noop {} {\enquote {\bibinfo {title} {Cosmic explorer strain sensitivity},}\ }\bibinfo {howpublished} {\url{https://dcc.cosmicexplorer.org/CE-T2000017/public}} (\bibinfo {year} {2023}),\ \bibinfo {note} {accessed: 07-03-2023}\BibitemShut {NoStop}%
\bibitem [{\citenamefont {Gupta}\ \emph {et~al.}(2023)\citenamefont {Gupta} \emph {et~al.}}]{Gupta:2023lga}%
  \BibitemOpen
  \bibfield  {author} {\bibinfo {author} {\bibfnamefont {Ish}\ \bibnamefont {Gupta}} \emph {et~al.},\ }\bibfield  {title} {\enquote {\bibinfo {title} {{Characterizing Gravitational Wave Detector Networks: From A$^\sharp$ to Cosmic Explorer}},}\ }\href@noop {} {\  (\bibinfo {year} {2023})},\ \Eprint {http://arxiv.org/abs/2307.10421} {arXiv:2307.10421 [gr-qc]} \BibitemShut {NoStop}%
\bibitem [{\citenamefont {LIGO~Scientific}(2022)}]{post-o5-curves-obs-paper}%
  \BibitemOpen
  \bibfield  {author} {\bibinfo {author} {\bibfnamefont {Virgo}\ \bibnamefont {LIGO~Scientific}},\ }\href@noop {} {\enquote {\bibinfo {title} {{Noise curves used for Simulations in the update of the Observing Scenarios Paper}},}\ }\bibinfo {howpublished} {\url{https://dcc.ligo.org/LIGO-T2000012/public}} (\bibinfo {year} {2022}),\ \bibinfo {note} {[Accessed 04-01-2022]}\BibitemShut {NoStop}%
\bibitem [{\citenamefont {Collaboration}(2021)}]{et-noise-curve}%
  \BibitemOpen
  \bibfield  {author} {\bibinfo {author} {\bibfnamefont {ET}~\bibnamefont {Collaboration}},\ }\href@noop {} {\enquote {\bibinfo {title} {{ET sensitivities page }},}\ }\bibinfo {howpublished} {\url{https://www.et-gw.eu/index.php/etsensitivities#datafiles}} (\bibinfo {year} {2021}),\ \bibinfo {note} {[Accessed 04-11-2023]}\BibitemShut {NoStop}%
\bibitem [{\citenamefont {Hild}\ \emph {et~al.}(2008)\citenamefont {Hild}, \citenamefont {Chelkowski},\ and\ \citenamefont {Freise}}]{Hild:2008ng}%
  \BibitemOpen
  \bibfield  {author} {\bibinfo {author} {\bibfnamefont {Stefan}\ \bibnamefont {Hild}}, \bibinfo {author} {\bibfnamefont {Simon}\ \bibnamefont {Chelkowski}}, \ and\ \bibinfo {author} {\bibfnamefont {Andreas}\ \bibnamefont {Freise}},\ }\bibfield  {title} {\enquote {\bibinfo {title} {{Pushing towards the ET sensitivity using 'conventional' technology}},}\ }\href@noop {} {\  (\bibinfo {year} {2008})},\ \Eprint {http://arxiv.org/abs/0810.0604} {arXiv:0810.0604 [gr-qc]} \BibitemShut {NoStop}%
\bibitem [{\citenamefont {Hild}\ \emph {et~al.}(2010)\citenamefont {Hild}, \citenamefont {Chelkowski}, \citenamefont {Freise}, \citenamefont {Franc}, \citenamefont {Morgado}, \citenamefont {Flaminio},\ and\ \citenamefont {DeSalvo}}]{Hild:2009ns}%
  \BibitemOpen
  \bibfield  {author} {\bibinfo {author} {\bibfnamefont {Stefan}\ \bibnamefont {Hild}}, \bibinfo {author} {\bibfnamefont {Simon}\ \bibnamefont {Chelkowski}}, \bibinfo {author} {\bibfnamefont {Andreas}\ \bibnamefont {Freise}}, \bibinfo {author} {\bibfnamefont {Janyce}\ \bibnamefont {Franc}}, \bibinfo {author} {\bibfnamefont {Nazario}\ \bibnamefont {Morgado}}, \bibinfo {author} {\bibfnamefont {Raffaele}\ \bibnamefont {Flaminio}}, \ and\ \bibinfo {author} {\bibfnamefont {Riccardo}\ \bibnamefont {DeSalvo}},\ }\bibfield  {title} {\enquote {\bibinfo {title} {{A Xylophone Configuration for a third Generation Gravitational Wave Detector}},}\ }\href {\doibase 10.1088/0264-9381/27/1/015003} {\bibfield  {journal} {\bibinfo  {journal} {Class. Quant. Grav.}\ }\textbf {\bibinfo {volume} {27}},\ \bibinfo {pages} {015003} (\bibinfo {year} {2010})},\ \Eprint {http://arxiv.org/abs/0906.2655} {arXiv:0906.2655 [gr-qc]} \BibitemShut {NoStop}%
\bibitem [{\citenamefont {Hild}\ \emph {et~al.}(2011)\citenamefont {Hild} \emph {et~al.}}]{Hild:2010id}%
  \BibitemOpen
  \bibfield  {author} {\bibinfo {author} {\bibfnamefont {S.}~\bibnamefont {Hild}} \emph {et~al.},\ }\bibfield  {title} {\enquote {\bibinfo {title} {{Sensitivity Studies for Third-Generation Gravitational Wave Observatories}},}\ }\href {\doibase 10.1088/0264-9381/28/9/094013} {\bibfield  {journal} {\bibinfo  {journal} {Class. Quant. Grav.}\ }\textbf {\bibinfo {volume} {28}},\ \bibinfo {pages} {094013} (\bibinfo {year} {2011})},\ \Eprint {http://arxiv.org/abs/1012.0908} {arXiv:1012.0908 [gr-qc]} \BibitemShut {NoStop}%
\bibitem [{\citenamefont {Baibhav}\ \emph {et~al.}(2019)\citenamefont {Baibhav}, \citenamefont {Berti}, \citenamefont {Gerosa}, \citenamefont {Mapelli}, \citenamefont {Giacobbo}, \citenamefont {Bouffanais},\ and\ \citenamefont {Di~Carlo}}]{Baibhav:2019gxm}%
  \BibitemOpen
  \bibfield  {author} {\bibinfo {author} {\bibfnamefont {Vishal}\ \bibnamefont {Baibhav}}, \bibinfo {author} {\bibfnamefont {Emanuele}\ \bibnamefont {Berti}}, \bibinfo {author} {\bibfnamefont {Davide}\ \bibnamefont {Gerosa}}, \bibinfo {author} {\bibfnamefont {Michela}\ \bibnamefont {Mapelli}}, \bibinfo {author} {\bibfnamefont {Nicola}\ \bibnamefont {Giacobbo}}, \bibinfo {author} {\bibfnamefont {Yann}\ \bibnamefont {Bouffanais}}, \ and\ \bibinfo {author} {\bibfnamefont {Ugo~N.}\ \bibnamefont {Di~Carlo}},\ }\bibfield  {title} {\enquote {\bibinfo {title} {{Gravitational-wave detection rates for compact binaries formed in isolation: LIGO/Virgo O3 and beyond}},}\ }\href {\doibase 10.1103/PhysRevD.100.064060} {\bibfield  {journal} {\bibinfo  {journal} {Phys. Rev. D}\ }\textbf {\bibinfo {volume} {100}},\ \bibinfo {pages} {064060} (\bibinfo {year} {2019})},\ \Eprint {http://arxiv.org/abs/1906.04197} {arXiv:1906.04197 [gr-qc]} \BibitemShut {NoStop}%
\bibitem [{\citenamefont {Lower}\ \emph {et~al.}(2018)\citenamefont {Lower}, \citenamefont {Thrane}, \citenamefont {Lasky},\ and\ \citenamefont {Smith}}]{Lower:2018seu}%
  \BibitemOpen
  \bibfield  {author} {\bibinfo {author} {\bibfnamefont {Marcus~E.}\ \bibnamefont {Lower}}, \bibinfo {author} {\bibfnamefont {Eric}\ \bibnamefont {Thrane}}, \bibinfo {author} {\bibfnamefont {Paul~D.}\ \bibnamefont {Lasky}}, \ and\ \bibinfo {author} {\bibfnamefont {Rory}\ \bibnamefont {Smith}},\ }\bibfield  {title} {\enquote {\bibinfo {title} {{Measuring eccentricity in binary black hole inspirals with gravitational waves}},}\ }\href {\doibase 10.1103/PhysRevD.98.083028} {\bibfield  {journal} {\bibinfo  {journal} {Phys. Rev. D}\ }\textbf {\bibinfo {volume} {98}},\ \bibinfo {pages} {083028} (\bibinfo {year} {2018})},\ \Eprint {http://arxiv.org/abs/1806.05350} {arXiv:1806.05350 [astro-ph.HE]} \BibitemShut {NoStop}%
\bibitem [{\citenamefont {Favata}\ \emph {et~al.}(2022)\citenamefont {Favata}, \citenamefont {Kim}, \citenamefont {Arun}, \citenamefont {Kim},\ and\ \citenamefont {Lee}}]{Favata:2021vhw}%
  \BibitemOpen
  \bibfield  {author} {\bibinfo {author} {\bibfnamefont {Marc}\ \bibnamefont {Favata}}, \bibinfo {author} {\bibfnamefont {Chunglee}\ \bibnamefont {Kim}}, \bibinfo {author} {\bibfnamefont {K.~G.}\ \bibnamefont {Arun}}, \bibinfo {author} {\bibfnamefont {JeongCho}\ \bibnamefont {Kim}}, \ and\ \bibinfo {author} {\bibfnamefont {Hyung~Won}\ \bibnamefont {Lee}},\ }\bibfield  {title} {\enquote {\bibinfo {title} {{Constraining the orbital eccentricity of inspiralling compact binary systems with Advanced LIGO}},}\ }\href {\doibase 10.1103/PhysRevD.105.023003} {\bibfield  {journal} {\bibinfo  {journal} {Phys. Rev. D}\ }\textbf {\bibinfo {volume} {105}},\ \bibinfo {pages} {023003} (\bibinfo {year} {2022})},\ \Eprint {http://arxiv.org/abs/2108.05861} {arXiv:2108.05861 [gr-qc]} \BibitemShut {NoStop}%
\bibitem [{\citenamefont {Pankow}\ \emph {et~al.}(2018)\citenamefont {Pankow} \emph {et~al.}}]{Pankow:2018qpo}%
  \BibitemOpen
  \bibfield  {author} {\bibinfo {author} {\bibfnamefont {Chris}\ \bibnamefont {Pankow}} \emph {et~al.},\ }\bibfield  {title} {\enquote {\bibinfo {title} {{Mitigation of the instrumental noise transient in gravitational-wave data surrounding GW170817}},}\ }\href {\doibase 10.1103/PhysRevD.98.084016} {\bibfield  {journal} {\bibinfo  {journal} {Phys. Rev. D}\ }\textbf {\bibinfo {volume} {98}},\ \bibinfo {pages} {084016} (\bibinfo {year} {2018})},\ \Eprint {http://arxiv.org/abs/1808.03619} {arXiv:1808.03619 [gr-qc]} \BibitemShut {NoStop}%
\bibitem [{\citenamefont {Baird}\ \emph {et~al.}(2013)\citenamefont {Baird}, \citenamefont {Fairhurst}, \citenamefont {Hannam},\ and\ \citenamefont {Murphy}}]{Baird:2012cu}%
  \BibitemOpen
  \bibfield  {author} {\bibinfo {author} {\bibfnamefont {Emily}\ \bibnamefont {Baird}}, \bibinfo {author} {\bibfnamefont {Stephen}\ \bibnamefont {Fairhurst}}, \bibinfo {author} {\bibfnamefont {Mark}\ \bibnamefont {Hannam}}, \ and\ \bibinfo {author} {\bibfnamefont {Patricia}\ \bibnamefont {Murphy}},\ }\bibfield  {title} {\enquote {\bibinfo {title} {{Degeneracy between mass and spin in black-hole-binary waveforms}},}\ }\href {\doibase 10.1103/PhysRevD.87.024035} {\bibfield  {journal} {\bibinfo  {journal} {Phys. Rev. D}\ }\textbf {\bibinfo {volume} {87}},\ \bibinfo {pages} {024035} (\bibinfo {year} {2013})},\ \Eprint {http://arxiv.org/abs/1211.0546} {arXiv:1211.0546 [gr-qc]} \BibitemShut {NoStop}%
\bibitem [{\citenamefont {Skilling}(2006)}]{Skilling:2006gxv}%
  \BibitemOpen
  \bibfield  {author} {\bibinfo {author} {\bibfnamefont {John}\ \bibnamefont {Skilling}},\ }\bibfield  {title} {\enquote {\bibinfo {title} {{Nested sampling for general Bayesian computation}},}\ }\href {\doibase 10.1214/06-BA127} {\bibfield  {journal} {\bibinfo  {journal} {Bayesian Analysis}\ }\textbf {\bibinfo {volume} {1}},\ \bibinfo {pages} {833--859} (\bibinfo {year} {2006})}\BibitemShut {NoStop}%
\bibitem [{\citenamefont {Higson}\ \emph {et~al.}(2018)\citenamefont {Higson}, \citenamefont {Handley}, \citenamefont {Hobson},\ and\ \citenamefont {Lasenby}}]{Higson2018}%
  \BibitemOpen
  \bibfield  {author} {\bibinfo {author} {\bibfnamefont {Edward}\ \bibnamefont {Higson}}, \bibinfo {author} {\bibfnamefont {Will}\ \bibnamefont {Handley}}, \bibinfo {author} {\bibfnamefont {Michael}\ \bibnamefont {Hobson}}, \ and\ \bibinfo {author} {\bibfnamefont {Anthony}\ \bibnamefont {Lasenby}},\ }\bibfield  {title} {\enquote {\bibinfo {title} {Dynamic nested sampling: an improved algorithm for parameter estimation and evidence calculation},}\ }\href {\doibase 10.1007/s11222-018-9844-0} {\bibfield  {journal} {\bibinfo  {journal} {Statistics and Computing}\ }\textbf {\bibinfo {volume} {29}},\ \bibinfo {pages} {891–913} (\bibinfo {year} {2018})}\BibitemShut {NoStop}%
\bibitem [{\citenamefont {Chen}\ and\ \citenamefont {Shao}(1999)}]{Chen1999}%
  \BibitemOpen
  \bibfield  {author} {\bibinfo {author} {\bibfnamefont {Ming-Hui}\ \bibnamefont {Chen}}\ and\ \bibinfo {author} {\bibfnamefont {Qi-Man}\ \bibnamefont {Shao}},\ }\bibfield  {title} {\enquote {\bibinfo {title} {Monte carlo estimation of bayesian credible and hpd intervals},}\ }\href {\doibase 10.1080/10618600.1999.10474802} {\bibfield  {journal} {\bibinfo  {journal} {Journal of Computational and Graphical Statistics}\ }\textbf {\bibinfo {volume} {8}},\ \bibinfo {pages} {69–92} (\bibinfo {year} {1999})}\BibitemShut {NoStop}%
\bibitem [{\citenamefont {Dhurkunde}\ \emph {et~al.}(2022)\citenamefont {Dhurkunde}, \citenamefont {Fehrmann},\ and\ \citenamefont {Nitz}}]{Dhurkunde:2021csz}%
  \BibitemOpen
  \bibfield  {author} {\bibinfo {author} {\bibfnamefont {Rahul}\ \bibnamefont {Dhurkunde}}, \bibinfo {author} {\bibfnamefont {Henning}\ \bibnamefont {Fehrmann}}, \ and\ \bibinfo {author} {\bibfnamefont {Alexander~H.}\ \bibnamefont {Nitz}},\ }\bibfield  {title} {\enquote {\bibinfo {title} {{Hierarchical approach to matched filtering using a reduced basis}},}\ }\href {\doibase 10.1103/PhysRevD.105.103001} {\bibfield  {journal} {\bibinfo  {journal} {Phys. Rev. D}\ }\textbf {\bibinfo {volume} {105}},\ \bibinfo {pages} {103001} (\bibinfo {year} {2022})},\ \Eprint {http://arxiv.org/abs/2110.13115} {arXiv:2110.13115 [astro-ph.IM]} \BibitemShut {NoStop}%
\bibitem [{\citenamefont {Soni}\ \emph {et~al.}(2024)\citenamefont {Soni}, \citenamefont {Dhurandhar},\ and\ \citenamefont {Mitra}}]{Soni:2023veu}%
  \BibitemOpen
  \bibfield  {author} {\bibinfo {author} {\bibfnamefont {Kanchan}\ \bibnamefont {Soni}}, \bibinfo {author} {\bibfnamefont {Sanjeev}\ \bibnamefont {Dhurandhar}}, \ and\ \bibinfo {author} {\bibfnamefont {Sanjit}\ \bibnamefont {Mitra}},\ }\bibfield  {title} {\enquote {\bibinfo {title} {{Obtaining statistical significance of gravitational wave signals in hierarchical search}},}\ }\href {\doibase 10.1103/PhysRevD.109.024046} {\bibfield  {journal} {\bibinfo  {journal} {Phys. Rev. D}\ }\textbf {\bibinfo {volume} {109}},\ \bibinfo {pages} {024046} (\bibinfo {year} {2024})},\ \Eprint {http://arxiv.org/abs/2309.00019} {arXiv:2309.00019 [astro-ph.IM]} \BibitemShut {NoStop}%
\bibitem [{\citenamefont {Hobbs}\ \emph {et~al.}(2005)\citenamefont {Hobbs}, \citenamefont {Lorimer}, \citenamefont {Lyne},\ and\ \citenamefont {Kramer}}]{Hobbs:2005yx}%
  \BibitemOpen
  \bibfield  {author} {\bibinfo {author} {\bibfnamefont {George}\ \bibnamefont {Hobbs}}, \bibinfo {author} {\bibfnamefont {D.~R.}\ \bibnamefont {Lorimer}}, \bibinfo {author} {\bibfnamefont {A.~G.}\ \bibnamefont {Lyne}}, \ and\ \bibinfo {author} {\bibfnamefont {M.}~\bibnamefont {Kramer}},\ }\bibfield  {title} {\enquote {\bibinfo {title} {{A Statistical study of 233 pulsar proper motions}},}\ }\href {\doibase 10.1111/j.1365-2966.2005.09087.x} {\bibfield  {journal} {\bibinfo  {journal} {Mon. Not. Roy. Astron. Soc.}\ }\textbf {\bibinfo {volume} {360}},\ \bibinfo {pages} {974--992} (\bibinfo {year} {2005})},\ \Eprint {http://arxiv.org/abs/astro-ph/0504584} {arXiv:astro-ph/0504584} \BibitemShut {NoStop}%
\end{thebibliography}%
\end{document}